\documentclass[article]{aa}
\usepackage{graphicx}
\usepackage{aalongtable}
\usepackage{lscape}
\usepackage{txfonts}
\usepackage{natbib}
\bibpunct{(}{)}{;}{a}{}{,}
\usepackage{color}
\newcommand{\sext}{\texttt{SExtractor}}

\newcommand{\kf}{$\rm{K}$}
\newcommand{\jf}{$\rm{J}$}
\newcommand{\dev}{$r^{1/4}$}
\newcommand{\fwhm}{\texttt{FWHM}}
\newcommand{\mumax}{\ensuremath{\mu_{\rm{max}}}}

\newcommand{\ie}{{\em i.e.}}
\newcommand{\eg}{{\em e.g.}}
\newcommand{\rb}[1]{\raisebox{1.3ex}[0pt]{#1}}

\begin{document}
\title{WINGS: a WIde-field Nearby Galaxy-cluster Survey.}

\subtitle{III. Deep near-infrared photometry of 28 nearby clusters
\thanks{Based on observations taken at the United Kingdom Infra-Red 
Telescope, operated by the  Joint Astronomy  Centre  on behalf of  the
Science and Technology Facilities Council of the U.K.}}

   \author{
 T. Valentinuzzi\inst{1} \and
 D. Woods\inst{2,3} \and
 G. Fasano\inst{4} \and
 M. Riello\inst{5} \and
 M. D'Onofrio\inst{1} \and
 J. Varela\inst{4} \and
 D. Bettoni\inst{4} \and
 A. Cava\inst{4,6}\and
 W.J. Couch\inst{7}\and
 A. Dressler\inst{8}\and
 J. Fritz\inst{4} \and
 M. Moles\inst{9} \and
 A. Omizzolo\inst{10,4} \and
 B.M. Poggianti\inst{4} \and
 P. Kj{\ae}rgaard\inst{11}
          }

   \institute{
     Astronomy Department, University of Padova, Vicolo Osservatorio 3, 35122 Padova, Italy \and
     School of Physics, University of New South Wales, Sydney 2052, NSW, Australia \and
     Dept. of Physics \& Astronomy, University of British Columbia, 6224 Agricultural Road, Vancouver, V6T 1Z1, BC, Canada \and
     INAF -- Padova Astronomical Observatory, Vicolo Osservatorio 5, 35122 Padova, Italy \and
     Institute of Astronomy, Madingley Rd., Cambridge CB3 0HA \and
     Instituto de Astrofisica de Canarias, Via Lactea s.n., 38205 La Laguna, Tenerife, Spain \and
     Centre for Astrophysics \& Supercomputing, Swinburne University, Hawthorn 3122, VIC, Australia\and
     Observatories of the Carnegie Institution of Washington, Pasadena, CA 91101, USA\and
     Instituto de Astrof\'{\i}sica de Andaluc\'{\i}a (C.S.I.C.) Apartado 3004, 18080 Granada, Spain \and
     Specola Vaticana, 00120 Stato Citta' del Vaticano \and
     The Niels Bohr Institute, Juliane Maries Vej 30, 2100 Copenhagen, Denmark
    }

   \offprints{Tiziano Valentinuzzi, \email{tiziano.valentinuzzi@unipd.it}}
   \date{\today}

%\abstract{}{}{}{}{}
% 5 {} token are mandatory

\abstract
% context heading (optional) % leave it empty if  necessary
{This      is  the   third  paper   in    a   series devoted    to the
{\textit{WIde-field Nearby  Galaxy-cluster Survey}} (WINGS).  WINGS is
a long-term  project aimed at  gathering wide-field, multiband imaging
and  spectroscopy   of galaxies  in  a complete   sample of   77 X-ray
selected, nearby clusters ($0.04\!\!<\!\!z\!\!<\!\!0.07$) located  far
from the galactic plane  ($\left|b\right|\geq20^o$).  The main goal of
this project is to establish a local reference sample for evolutionary
studies of galaxies and galaxy clusters.}
{This paper presents the near-infrared (J,K) photometric catalogs of
28 clusters of the WINGS sample and describes the procedures followed
to construct them.}
{The  raw data has   been reduced at CASU and   special care has  been
devoted  to the  final  coadding,  drizzling  technique,   astrometric
solution, and magnitude  calibration  for the WFCAM  pipeline-processed
data. We constructed the  photometric catalogs based on the final
calibrated,  coadded mosaics  ($\approx\!0.79\;\rm{deg}^2$)  in  J  (19
clusters) and K (27 clusters) bands. A customized interactive pipeline
was used  to clean  the catalogs  and  to  make mock images   for
photometric errors and completeness estimates.}
{We       provide            deep       near-infrared      photometric
catalogs\footnote{Tab.\ref{tab:Catalogs}    with the  complete     NIR
photometric catalogs are only available in electronic  form at the CDS
via anonymous  ftp  to   cdsarc.u-strasbg.fr (130.79.128.5)  or    via
http://cdsweb.u-strasbg.fr/cgi-bin/qcat?J/A+A/} (90\%   complete    in
detection     rate   at    total     magnitudes    $J\!\approx\!20.5$,
$K\!\approx\!19.4$, and in  classification  rate at $J\!\approx\!19.5$
and  $K\!\approx\!18.5$),  giving  positions, geometrical  parameters,
total and aperture magnitudes   for all  detected sources.   For  each
field we classify the detected sources as stars, galaxies, and objects
of ``unknown'' nature.}
{}

\titlerunning{WINGS III: Near-Infrared Catalogs}
\maketitle

%
%________________________________________________________________

%%%%%%%%%%%%%%%%%%%%%%%%%%%%%%%%%%%%%%%%%%%%%%%%%%%%%%%%%%%%%%%%%%%%%%
%%%%%%%%%%%%%%%%%%%%%%%%%% INTRODUCTION %%%%%%%%%%%%%%%%%%%%%%%%%%%%%%
%%%%%%%%%%%%%%%%%%%%%%%%%%%%%%%%%%%%%%%%%%%%%%%%%%%%%%%%%%%%%%%%%%%%%%

\section{Introduction}

Clusters of galaxies are important for studies of galaxy formation and
evolution, because they contain  a {\it volume-limited}  population of
galaxies observed   {\it   at the same   cosmic   epoch}.  Galaxies in
clusters are known   to follow tight  color-magnitude relations, which
appear to be universal and to have very small intrinsic scatter to the
highest                  redshifts               yet          observed
\citep[see, amongst others,][]{blakeslee03,holden04,mei06,delucia07}.   Together
with the  conventional interpretation of  the color-magnitude relation
as a  mass-metallicity   correlation \citep[\ie,][]{trager00},    this
implies that  the  majority of the  stellar  populations in early-type
cluster galaxies  were  formed  via  rapid  dissipative starbursts  at
$z\!>\!2$.   Fundamental-plane    studies  of   high-redshift  cluster
galaxies also support this  conclusion, at least  for the more massive
objects \citep[\ie,][]{vandokkum03,holden05},  although  the low  mass
galaxies seem to have undergone more extended star formation histories
\citep[\ie,][]{poggianti01}.

Theoretically, the existence of such  massive and old galaxies at high
redshift should represent a  strong challenge to models where galaxies
are assembled hierarchically,  from a  sequence  of major mergers   at
progressively lower redshifts. It is not possible, however, to exclude
by   spectrophotometry  alone,  these   galaxies  being assembled from
sub-units whose star formation has  already ceased, but which are  not
accreted  until later (the  so-called `dry' mergers).  This is assumed
to  be the main channel by  which spheroids grow   at $z\!<\!1$ in the
hierarchical scenario.

On  the other hand, if   galaxies are formed   via mergers, we  should
observe a steady decrease in the  mean stellar mass  in galaxies as we
go to earlier cosmic times and the most  massive members of the merger
tree would branch into even smaller twigs
\citep[][]{delucia06}. While it   is  generally difficult   to measure
galaxy masses, the \kf-band luminosity function is believed to provide
an adequate surrogate
\citep[][]{kauffmann98}. Indeed, it is known that the rest frame
$\rm{H}$ or \kf\ luminosity of galaxies is seen to correlate well with
stellar  and even dynamical mass  for local and high-redshift galaxies
\citep[see, \ie,][]{kodama03}.

To  reach higher and  higher   redshifts  it  is surely necessary   to
adequately  constrain galaxy   formation models;  on   the other hand,
precise  knowledge of the  properties of clusters and cluster galaxies
in the local Universe is necessary, as a benchmark for higher redshift
studies. The \textit{WIde-field Nearby Galaxy-clusters Survey}
\citep[][hereafter  Paper-I]{wingsI06} is an answer to this
need. WINGS\footnote{Please refer to WINGS Website for updated details
on the survey and its products, \texttt{http://web.oapd.inaf.it/wings}
} is a long-term multiwavelength survey specially designed to provide
the first robust characterization of both the photometric and
spectroscopic properties of galaxies in nearby clusters, as well as
determine the variations in these properties.

The   survey  core, based   on  optical  B,V  imaging  of  77   nearby
($\langle\rm{z}\rangle\!\approx\!0.05$) galaxy-clusters
\citep[see,][hereafter Paper-II]{wingsII08},  has been complemented by
several ancillary  projects:  (i) a   spectroscopic  follow up  of   a
subsample of 51 clusters,  obtained with the spectrographs  WYFFOS@WHT
and 2dF@AAT;  (ii) near-infrared (J,K)   imaging of a  subsample of 28
clusters obtained with WFCAM@UKIRT, presented here; (iii) U broad- and
H$_\alpha$  narrow-band  imaging   of subsamples  of   WINGS clusters,
obtained with  wide-field cameras at  different  telescopes (INT, LBT,
Bok).  The observations and data reduction for the first two follow-up
projects have been completed
\citep[][in  preparation; this paper]{cava08,fritz09},  while the observations
for the H$_\alpha$ and U-band surveys are still ongoing.

The near  infrared  section  of the  WINGS  survey  (WINGS-NIR)   is a
collection of $\approx\!0.79\;\rm{deg}^2$ mosaics   in (J,K) bands of  28
nearby clusters (17 clusters have been observed in both bands).

WINGS-NIR  is by far the largest  survey of  nearby galaxy clusters as
far as  the  area coverage is  concerned.  In  fact,  in this redshift
range, only  individual  clusters or small  cluster samples  have been
studied       in       the         literature     up      to       now
\citep[\eg,][]{pahre99,gavazzi90,depropris03}. The WINGS    survey  of
near-infrared data consists   of nearly one  million detected sources,
with   150,000   and 500,000  well     classified stars  and galaxies,
respectively. 

In section   II of this     paper we describe observations   and  data
reduction   techniques, including a  brief   presentation of both  the
instrumentation and  the software pipeline   used to create  the final
coadded mosaics.   Section III presents  the step-by-step procedure of
the catalog  production  pipeline, including   an extensive discussion
about star/galaxy  classification  and  interactive  cleaning  of  the
catalogs.   Section IV deals with  data quality assessment and overall
properties of   the  catalogs,  like  completeness,  astrometric   and
photometric accuracy and precision.  In the last  section we present a
brief summary of the main features of WINGS-NIR survey and catalogs.

Throughout  this    paper we  will  use     a cosmological  model with
parameters:         \mbox{$H_0=70$       km$\,$s$^{-1}\,$Mpc$^{-1}$},
$\Omega_M=0.3$ and $\Omega_\Lambda=0.7$.

%%%%%%%%%%%%%%%%%%%%%%%%%%%%%%%%%%%%%%%%%%%%%%%%%%%%%%%%%%%%%%%%%%%%%%
%%%%%%%%%%%%%%%%%%%%%%%%%%%% THE DATA %%%%%%%%%%%%%%%%%%%%%%%%%%%%%%%%
%%%%%%%%%%%%%%%%%%%%%%%%%%%%%%%%%%%%%%%%%%%%%%%%%%%%%%%%%%%%%%%%%%%%%%

\section{Observations and Data Reduction}\label{sec:data}

\begin{figure}
\centering            
\includegraphics[scale=0.5,bb=5.3cm 6cm 17cm 14cm ]{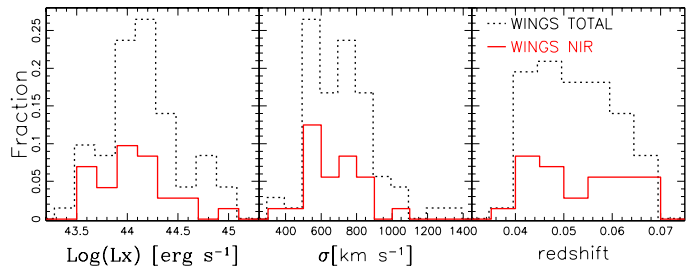}
\caption{Distribution of WINGS-NIR clusters (red full line) compared with the
distribution of all the WINGS survey clusters (black dotted line). The
dotted diagram is slightly shifted for ease of view.
\label{fig:sampling}}
\end{figure}

The near-infrared  data  have been collected  at   the UKIRT telescope
using   the WFCAM   instrument   \citep[see,][]{casali07} during  four
observing semesters,   from   April   2005  to  April    2007   (see
Tab.\ref{tab:runs}). The   images  where  obtained  in the  J   and  K
broad-bands of  the  Mauna Kea  photometric system\footnote{Note  that
WFCAM is equipped with a broad K-band filter,  not with the short-band
$\rm{K_{s}}$.}
\citep[][]{Tokunaga02}.
 The original plan  of the WINGS-NIR survey was  to image at least all
 the  WINGS clusters with spectroscopy   and visible imaging by UKIRT:
 unfortunately bad weather (see Tab.\ref{tab:runs}) limited the sample
 to   only 28   clusters.     However,   as  it  can    be   seen   in
 Fig.\ref{fig:sampling} the NIR subsample  has good coverage  of X-ray
 luminosities and redshifts when compared with the total WINGS sample,
 while   it is  slightly  biased   towards the   low  cluster velocity
 dispersion regime.

\begin{table*}
\begin{center}
\caption{Summary of the observations, taken with WFCAM on UKIRT,
discussed  in this paper. 
\label{tab:runs}}
\begin{tabular}{cccccc}
%\hline
 Semester &PATT Ref. & Alloc. & Completion & Observation Dates & Comments \\
\hline
\hline
       &         &     &      & Apr29-May 4         & Compromised by poor   \\	
 2005A & U/05A/2 & 40h & 67\% & May 14-15(Pq)       & weather, seeing and   \\
       &         &     &      & May 28-29(Pq), 2005 & camera
misalignment.  \\
\hline
       &          &     &      & Sept 30              & Severely  \\
 2005B & U/05B/12 & 40h & 18\% & Nov 27, 2005         & impacted by \\ 
       &          &     &      & Jan 6, 2006 (all Pq) & poor weather. \\
\hline
       &          &     &       & May 20\&26(Pq)    & Good\\
 2006A & U/06A/12 & 10h & 100\% & June 2(Pq)        & observing  \\
       &          &     &       & June 21\&23, 2006 & conditions.\\
\hline
       &          &     &      & Dec. 14, 2006 & \\
 2006B & U/06B/12 & 14h & 65\% & March 20, 28, 31 & Average conditions. \\
       &          &     &      & April 3-5, 10-11, 2007 (all Pq) & \\ 
\hline 
\end{tabular}
\end{center}
\center{Note  "(Pq)" refers to observations taken in the PATT queue, and not by a WINGS project observer.}
\end{table*}

\begin{figure*}
\centering            
\includegraphics[scale=1]{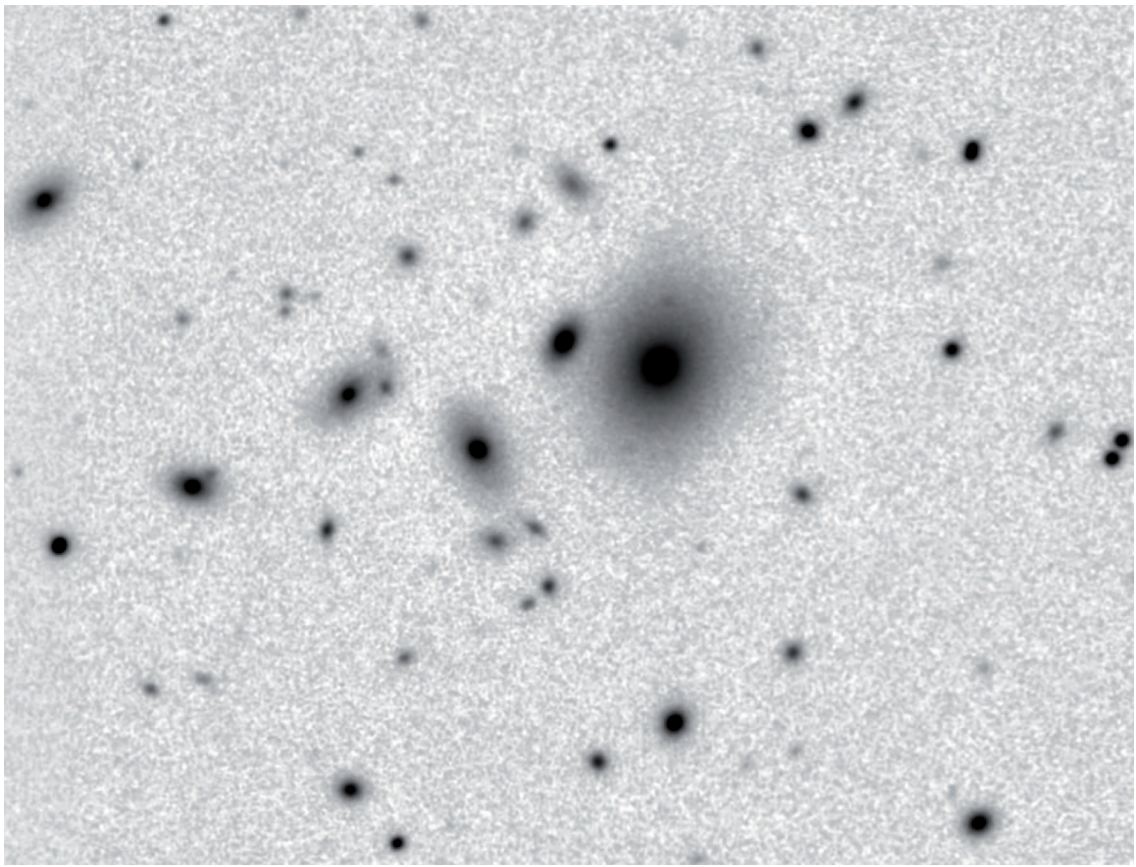}
\caption{The central region (2.5\arcmin x 2.2\arcmin, north is up, east is left) of the cluster A2124 K-band mosaic.
\label{fig:mosaic}}
\end{figure*}

\begin{figure}
\centering            
\includegraphics[scale=0.45]{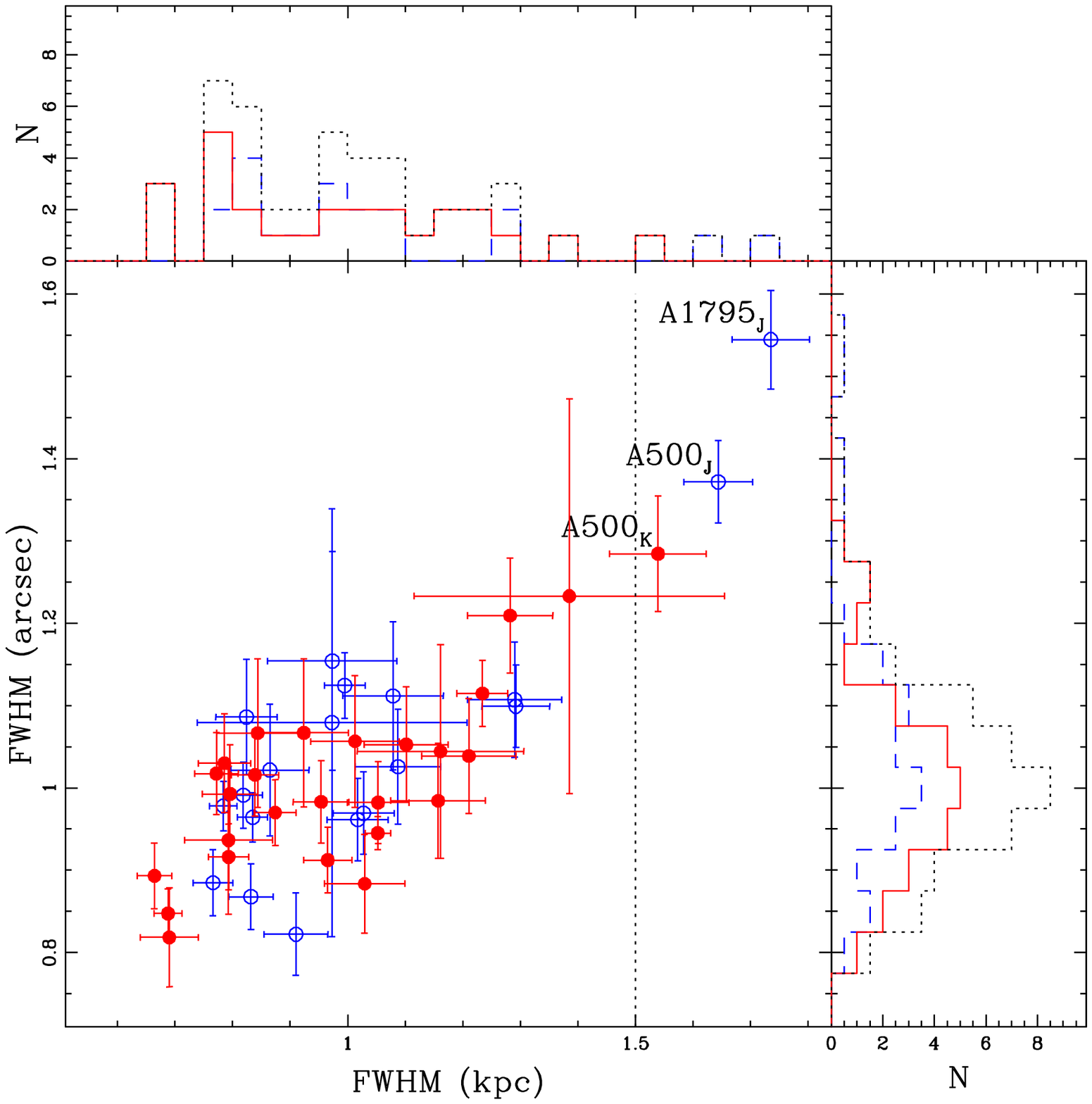}
\caption{The distribution of the median \fwhm s (angular vs. physical)
of all the J- (open blue circle;  blue dashed line) and K-band (filled
red circles; red  full line) mosaics: black dotted lines are the
total  J and K  distributions.   The error bars  show  the RMS of the
median seeing variation  of each  section  of the mosaic,  as shown in
Fig.\ref{fig:seeingvariation}. Note that a large error bar means that,
due to   non-photometric nights and/or   technical problems  with  the
camera, the  \fwhm\ was unstable  among the different exposures making
up the final mosaic. The dotted vertical line is the limit for quality
requirements  of WINGS survey  spatial physical resolution, needed
for reliable morphological classification and surface photometry.
\label{fig:fwhm}}
\end{figure}

WFCAM   is  an  assembly  of   four Rockwell  Hawaii-II  $2048$x$2048$
$18\mu$m-pixel  array detectors with 0\farcs4  pixel  size and a total
field of view of $0.21\,\rm{deg}^2$.  The four detectors are spaced at
$94\%$ of their width, at the corners of a field of about $0.4
\,\rm{deg}^2$. More details about the camera, imaging properties, defects
and drawbacks can be found in \citet{dye06}.  Fig.\ref{fig:fwhm} shows
the angular  vs.  physical median \fwhm  s for  J- (blue open circles;
dashed line) and K-band (red full circles; full line). It is seen that
more  than 50\% of our  images  have seeing below  1\farcs0, and  only
three out of 47 co-added mosaics do not  meet the quality requirements
of the  WINGS survey    regarding physical  size  resolution   (dotted
vertical line; see Paper-I).   This requirement is needed for reliable
morphological classification and surface  photometry.  A complete  and
useful  log   of  the  observations    concerning sky,   zero    point
fluctuations, etc. can be found in the CASU website\footnote{
\texttt{http://casu.ast.cam.ac.uk/surveys-projects/wfcam/data\-processing}}.

All WFCAM data, including proprietary  PI data, are pipeline-processed
by the Cambridge Astronomical Survey Unit (CASU) as  part of the VISTA
Data          Flow               System    (VDFS)          development
\citep[see,][for  an overview]{emerson04,irwin04,hambly04}. The data
products   of the  CASU   pipeline include  artifact-corrected images,
interleaved and stacked as appropriate, confidence maps and catalogues
of  detected  objects (both stellar  and  extended).  All  of them are
astrometrically and photometrically  calibrated with respect to 2MASS.
We  refer the reader to  \citet{irwin08} for a detailed description of
the pipeline and
\citet{hodgkin08} for a detailed description and discussion of the
photometric  calibration. An overview of  the pipeline and photometric
calibration can  also be found in  the  WFCAM section of  the CASU web
site~\footnote{The                     web                        page
\texttt{http://www.ast.cam.ac.uk/$\sim$wfcam} provides access  to  the
WFCAM raw data and  includes detailed description and useful summaries
of  the  observing and   reduction processes.},  which  contains other
quality control plots too, showing, for  each night, the variations of
the   photometric  zero-points, the  image    seeing  (\fwhm) and  sky
background  level  \citep[see][for an overview]{riello08}  that proved
very useful for  assessing the quality of  the data  presented in this
paper.

\begin{figure}
  \centering \includegraphics[scale=0.25]{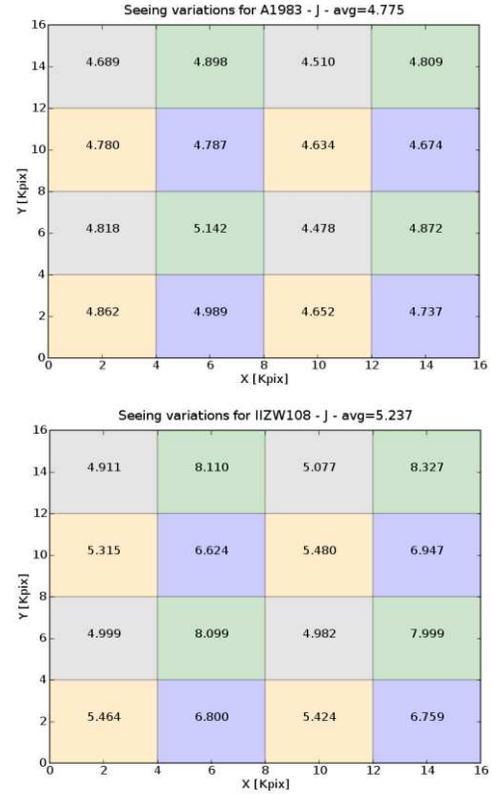}
  \caption {Due  to  the    WFCAM  configuration  and   the    way the
  observations  are carried out,  the   final mosaics have global  PSF
  variations in a chessboard pattern fashion.  The `tile' described in
  the  text, is composed  of 4 exposures  (\ie,  MEFs), of 4 detectors
  each, which  gives the 16 sections  of the mosaic shown  here (pixel
  size is 0\farcs2).  The best and the worst global PSF of our mosaics
  is  presented in the  top and the bottom   panel, respectively.  For
  each  of the $16$  sections, the  median  \fwhm\ (in pixels) of well
  classified   and non saturated stars    is reported. 
  \label{fig:seeingvariation}}
\end{figure}

\subsection{Pipeline data products}\label{sec:pipedp}

WFCAM observations are organised in groups, where  a group is composed
of  all the images taken with  a given filter, dither and, optionally,
micro-step  sequence.  For all  the clusters  presented in this paper,
observations were  taken using a    9-point dithering pattern with   a
$2\!\times\!2$ micro-step sequence  to partially  recover the  spatial
resolution~\footnote{Micro-stepping and   interlacing allow scales  of
$1/2$ and  $1/3$ of   the  original  pixel   size, avoiding  the   PSF
under-sampling  when   good seeing   conditions  occur.   In  all  our
observations the effective pixel size is 0\farcs2.}
\citep[][]{casali07}. Micro-stepping is done by shifting the
telescope by $n+\frac{1}{2}$  pixels (for  the $2\!\times\!2$ pattern)
and interleaving consists of an algorithm that creates an output image
that is a regular interwoven  pattern of all the  input pixels.   Some
relevant  caveats   about  interleaving   for   this  paper  are:   i)
interleaving  does not eliminate bad pixels,  ii) the PSF often varies
on short enough timescales  to lead to unusually  ``spiky" interleaved
PSFs that required additional treatment to deal with.  The first issue
was  addressed   by  adopting a  dithering  strategy   that  allows an
efficient rejection   of cosmic rays  and bad  pixels at  the stacking
phase. We will discuss the second issue in the following section.

Each  group is  therefore  composed of 36 independent  Multi-Extension
FITS files (MEF) with four  extensions (one per detector).  Given that
the WFCAM    four detectors are  separated,    in both  directions, by
$\approx\!90\%$  of their  size, it  is  necessary to take four images
(\ie, four   MEFs) with  appropriate  offset in    order to  survey  a
contiguous area  of     about   $0.79\,\rm{deg}^2$   (a  tile,     see
Fig.\ref{fig:seeingvariation}).  Depending  on  the  cluster  and  the
filter,  one or more  tiles were observed to achieve the desired
magnitude limit.

For each image  group the pipeline produces a  stacked image  from the
interleaved  (0\farcs2 sampled)   frames  coming from  each  dithering
point.    Each stacked  frame  is astrometrically  and photometrically
calibrated and has an accompanying object catalogue and confidence map
generated  for it.  These  data products  represent the  input for the
final stage of the processing which involves:

\begin{enumerate}
\item
  Stacking,  for each of the four  tile positions,  all the individual
  pipeline  stacks  (typically  2 for  J-band observations  and  4 for
  K-band ones). To  avoid confusion we  will refer to  these images as
  the (four) {\em final stacks}.
\item
  Mosaicking  the  four  final stacks  into  a  single  image covering
  $\approx\!0.79$ deg$^2$.
\end{enumerate}

\subsection{Final Stacks and Mosaics}\label{stacks}

The final stacks are produced using a stand-alone version of the image
stacking engine used by the CASU pipeline.  The main difference of the
stand-alone version is that it uses the  object catalogues for further
refinement of  the WCS  (World  Coordinate  System)  offsets  that are
stored in the FITS headers. In particular:

\begin{enumerate}

\item 
  The alignment of the input images is driven by the WCS of the input
  frames and then is further refined using the associated object
  catalogues. The first FITS image in the stacking list is used as a
  reference and the other images are resampled (using
  nearest-neighbour interpolation) onto the WCS of the first image.
\item 
  All the images are then scaled at the detector level, using additive
  corrections, to match the background of the common overlap area.
\item 
  The next step  involves   rejecting  bad  pixels using  the    input
  confidence maps and an iterative $k$-sigma  clipping that uses a sky
  background-calibrated Poisson noise model  for the frame (which uses
  a robust   MAD\footnote{MAD:  median   absolute deviation   from the
  median.}-based  sky noise  estimator to  define  the equivalent
  RMS noise level).
\item 
  Finally,   pixels  are  combined  using  inverse variance  weighting
  derived from the  combination of the  input confidence maps  and the
  average noise properties  of the image  with an additional weighting
  according to the seeing (i.e. weight $\propto seeing^{-2}$).

\end{enumerate}

A two-dimensional non-linear iterative filter was then applied to
remove the background from the final stacks.  This additional step is
required because the stacked frames produced by the pipeline, although
corrected for residual reset-anomaly, gain and low-level sky
variations, still show background variations
($\sigma_{BG}=0.7\div1.2$ ADUs).  Finally we used a variant of
the drizzle algorithm \citep[see \eg,][]{fruchter02} to remap the
original 0\farcs4 pixels that were interleaved into a finer 0\farcs2
grid to obtain smoother PSFs (see Sec. \ref{sec:pipedp}).

The final stacks  are eventually combined  together to form a  mosaic,
again  using   a stand-alone version   of  the  pipeline software. The
software uses the photometric zero points  associated with each of the
four input images to match them  photometrically and then project them
on  the  same grid   using a drizzle-like    algorithm. To reduce  the
artifacts caused by the poorer image edges, all  the input pixels with
an  associated confidence  lower than  $50\%$  were  rejected from the
final  mosaic. Finally  the mosaic  zero-points  are re-measured  with
respect to 2MASS using the same software used by the CASU pipeline
\citep[see][]{hodgkin08}, and a final confidence map for the mosaic is
produced from the confidence maps of the single stacks.

%%%%%%%%%%%%%%%%%%%%%%%%%%%%%%%%%%%%%%%%%%%%%%%%%%%%%%%%%%%%%%%%%%%%%%
%%%%%%%%%%%%%%%%%%%%% PHOTOMETRIC CATALOGS %%%%%%%%%%%%%%%%%%%%%%%%%%%
%%%%%%%%%%%%%%%%%%%%%%%%%%%%%%%%%%%%%%%%%%%%%%%%%%%%%%%%%%%%%%%%%%%%%%

\section{The Photometric Catalogs\label{sec:preprocess}}

After   preparing   the  final  photometrically   and  astrometrically
calibrated co-added mosaics (hereafter simply  called mosaics) we used
a  custom  pipeline of miscellaneous   scripts to generate  the source
lists.  We  will use \sext's \texttt{MAG\_AUTO}  \citep[with the
default  input parameters,][see  manual  for details]{bertin96} as the
default magnitude throughout this  paper, unless explicitly otherwise
mentioned.

Hereafter we   present  a step-by-step  schematic  description  of the
catalogs production:

\begin{itemize}
\item 
  selection of a polygon enclosing  the reliable region of the mosaic,
  whose   total  area is  reported   in Tab.\ref{tab:clustertab}.   In
  practice,  the edges  of the  mosaic  are excluded.  We also exclude
  objects that are dissected  by, or are not  far enough away from the
  edges to ensure their photometry is unaffected;
\item 
  running the Terapix software WeightWatchers to generate a flag mask,
  to {\bf locate} the detections inside the good region and those with
  confidence $<\!70\%$ as deduced from the confidence maps of
the final mosaics: we remind the reader that pixels from the single
stacks with confidence levels below 50\% were not included in the
final mosaic  (see previous section);
\item 
  preliminary running of  \sext, to evaluate the
  \fwhm\ of non saturated stars (\texttt{CLASS\_STAR}$>0.95$), to choose
  the  detection thresholds, and  to quickly check  the quality of the
  mosaics (astrometry, photometry, number counts ...), as explained in
  section \ref{sec:stargal};
\item 
  extraction  of several background  stamps from the mosaic, useful to
  prepare a background   image for photometric errors   estimates
  with  simulations.  A collection  of  synthetic stars and  galaxies
  (30\% of  exponential disks) is separately   added to the background
  image,  in an attempt to  best reproduce  the \fwhm\ distribution of
  the  real  image.  Detection and   classification rates of stars and
  galaxies are computed separately (see Tab.~\ref{tab:clustertab}) and
  are also used to fine tune the  \sext\ input parameters (see section
  \ref{sec:stargal});
\item 
  final  running of \sext\,   with the adjusted  input  parameters and
  partitioning of the main output catalog into  the catalogs of stars,
  galaxies and unknown objects. The  latter step is achieved here just
  relying  upon  \sext's   \texttt{CLASS\_STAR},  which   ranges  from
  0~(galaxies) to 1~(stars), in   the following  way\footnote{In
  paper-II the galaxies value was $0.2$: the difference  is due to the
  intrinsic   differences between  near-infrared  and  optical images.
  These values are estimated  using  simulations which  reproduce  the
  peculiar characteristics of the mosaics.}:

  \begin{tabular}{ll}
    \textbf{Stars}    & $\texttt{CLASS\_STAR} \ge 0.8$  \\
    \textbf{Galaxies} & $\texttt{CLASS\_STAR} \le 0.35$  \\
    \textbf{Unknown}  & $0.35<\texttt{CLASS\_STAR} <0.8$; \\
  \end{tabular}
\item  
  since the \texttt{CLASS\_STAR} parameter  is not reliable  enough in
  all circumstances, in  particular  when the  mosaic  is  affected by
  strong seeing  variations, a further step is  necessary in  order to
  improve the assignment  of each object  to the right catalog and  to
  remove from  all of them the  spurious detections.  For this task we
  used  a  custom interactive tool  which  produces various diagnostic
  plots,   and  allows   us  to  distinguish   between stars  and
  galaxies; %(see for instance Figure~\ref{fig:cleaning});
\item 
  final ``visual'' (interactive) cleaning of the mosaic to correct any
  remaining  blatant error both  in  detection and classification.  In
  this phase the mosaic is displayed with the corresponding markers of
  stars  and galaxies and, by  visual inspection,  the corrections are
  made  directly on the image,  saved on disk and  then applied to the
  catalogs.
\end{itemize}

In the following sections we will describe  in more detail some of the
aforementioned steps taken as part of the catalogs production.

%%%%%%%%%%%%%%%%%%%%%%%%%%%%%%%%%%%%%%%%%%%%%%%%%%%%%%%%%%%%%%%%%%%%%%
%%%%%%%%%%%%%%%%%% DETECTION AND CLASSIFICATION %%%%%%%%%%%%%%%%%%%%%%
%%%%%%%%%%%%%%%%%%%%%%%%%%%%%%%%%%%%%%%%%%%%%%%%%%%%%%%%%%%%%%%%%%%%%%

\subsection{Source Detection and Star/Galaxy Classification\label{sec:stargal}}

The   source detection was performed  by  running  \sext\ on the final
mosaics, convolved by  2D gaussian filters of size  chosen to be equal
to the  median \fwhm.  In most cases,  the low level of the background
rms obtained  with  the  drizzling techniques  of  the  CASU pipeline,
allowed us  to  use a 1.5$\sigma$ clipping   and a minimum area of  20
adjacent pixels as threshold    parameters.  These  parameter   values
allowed us  to  simultaneously  obtain a   small number  of   spurious
detections  of stars and  galaxies, and  deep enough magnitude limits.
In  general,   for each  image, the   right  combination of threshold,
minimum area and filter size was chosen relying upon the expected
number   counts    of  galaxies/stars and    on serendipitous  visual
inspection  of  marked detections with    the  SAO-DS9 display   tool.
Spurious detections are typically  misclassified as galaxies by \sext,
and usually  result from local  background fluctuations  and spikes or
cross-talk of saturated stars.
  
The preliminary star/galaxy classification was done relying upon
\sext's stellarity index (\texttt{CLASS\_STAR}). The
 median \fwhm\ of  the image to be  processed (\texttt{SEEING\_\fwhm})
 is the chief parameter affecting the
\sext's star/galaxy separation algorithm. The sigma-clipping  and
filtering   detection  parameters   used  in  \sext\  are  also  quite
important, since background  fluctuations and PSF distortions near the
detection threshold  can introduce  uncertainties correlated with such
quantities.

\begin{figure}
  \centering       
\includegraphics[scale=0.35]{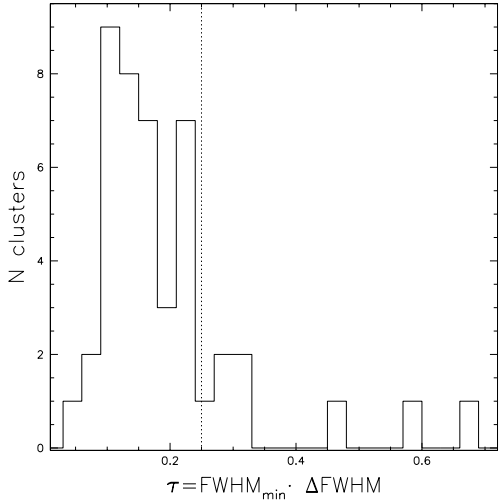}
  \caption     {Distribution     of        the     quantity
  $\tau=\fwhm_{\rm{MIN}}\cdot\Delta\fwhm$  of Eq.(\ref{eq:seeingrap})  for
  all the near-infrared   mosaics.    $\rm{\fwhm_{max/min}}$  are  the
  maximum and the minimum values   of the median \fwhm, calculated  on
  the single  detectors of   the final  stacked   tiles, as shown   in
  Fig.\ref{fig:seeingvariation}.     For       all     clusters   with
  $\tau\!\!<\!\!0.25$  (the  region  delimited by the  vertical dotted
  line) both the star/galaxy   classification and the cleaning  of the
  catalogs are quite straightforward.
\label{fig:seeingrap}}
\end{figure}

Aiming to test on our mosaics  both the detection capability of \sext\
and the reliability of  its  star/galaxy classifier, we have  produced
for each  cluster a mock image of  simulated stars and galaxies (a mix
of 70\% spheroidal \dev\ and of 30\% exponential  disks) with a sample
background coming from  the real image, and  trying to match as far as
possible the distribution of the stellar \fwhm.  With
\texttt{ARTDATA}  package,  synthetic stars were modeled  using Moffat
profiles with $\beta=2.5$.   For  spheroidal De~Vaucouleurs  and  disk
profiles,  intrinsic half-light radii  in   the range  1-5~kpc and  an
Euclidean power law (n=0.3) luminosity function were adopted.

\citet{foucaud07} claim that a 20\% increase of the value of
\texttt{SEEING\_\fwhm} parameter had to be used to correctly match the
synthetic stars added to the images, and this indicates the difficulty
\sext\ has to correctly classify stars, at fainter magnitudes, even in
good seeing conditions. We further  investigated this effect and found
that, while the \fwhm s  of stars produced by the  \texttt{ARTDATA-IRAF}
package are consistent  with the input  values, their \sext\ estimates
are $\approx$15\% higher. This drawback  can be dealt with following  the
\citet{foucaud07}  prescriptions   of    generating mock images   with
correspondingly diminished input values of the  \fwhm.  The results of
these simulations  are presented in Sec.~\ref{sec:completeness}, where
we discuss the completeness of the catalogs.

The      PSF     spatial     variation       exemplified   in
Fig.\ref{fig:seeingvariation}  is  by  far the   most important effect
which might complicate the correct choice of \texttt{CLASS\_STAR}. In
fact, while the  resulting \texttt{CLASS\_STAR} is highly sensitive to
the \texttt{SEEING\_\fwhm} keyword, up to  the last release of \sext\,
has this keyword fixed for the whole frame.   This means that, in case
of  strong  space  variations  of   the \fwhm,  \sext\  will  tend  to
overestimate  the number  of stars (galaxies)  in  the mosaic  regions
where the  local \fwhm\ is greater (lower)  than the median value used
for source extraction.

In Fig.\ref{fig:seeingrap} we  present, for  all our clusters  in
both bands, the distribution of the quantity:
\begin{equation}
\tau = \fwhm_{\rm{MIN}}\cdot(\fwhm_{\rm{MAX}}-\fwhm_{\rm{MIN}})  \label{eq:seeingrap}
\end{equation}
where  $\rm{\fwhm\_{max/min}}$ are the maximum  and the minimum median
\fwhm\ of the single stacked detectors,  respectively, as determined from
grids  like those in  Fig.\ref{fig:fwhm}.   The value  of $\tau$ is a
quality and  stability indicator of the  PSF  for complete mosaics and
all   the  observations    collected    during    a   given     night.
Fig.\ref{fig:seeingrap} shows that the majority  of our clusters  have
relatively low values (high quality)  of this quantity ($\tau<\!0.25$;
marked by the  vertical dotted line).   In clusters with $\tau>\!0.25$
the choice of the
\texttt{SEEING\_\fwhm} keyword and other parameters is very difficult,
due to  many misclassifications   being  found, and  the  use  of  the
interactive    cleaning   procedure  becomes  fundamental  to generate
reliable stars and galaxies catalogs (see Fig.\ref{fig:cleaning}).

%%%%%%%%%%%%%%%%%%%%%%%%%%%%%%%%%%%%%%%%%%%%%%%%%%%%%%%%%%%%%%%%%%%%%%
%%%%%%%%%%%%%%%%%%%%%% INTERACTIVE CLEANING %%%%%%%%%%%%%%%%%%%%%%%%%%
%%%%%%%%%%%%%%%%%%%%%%%%%%%%%%%%%%%%%%%%%%%%%%%%%%%%%%%%%%%%%%%%%%%%%%

\subsection{Interactive Cleaning \label{sec:cleaning}}

Spurious  detections are most frequently   found along the overlapping
regions of the  detectors, along the  mosaic edges (which are excluded
by the initial polygon mask) and in (usually  limited) regions of high
background fluctuations.  Cross-talk due to saturated stars is another
cause  of  spurious objects.  These false  detections can be  found at
fixed distances (in  symmetric  positions) from the saturated  objects
along  the  read-out   direction    of the  detector,  and    have   a
characteristic ``doughnut''  shape \citep[see, \ie,][]{dye06}.  It was
not possible to safely detect and delete  them automatically, as their
number and occurrence is not simply correlated with the peak intensity
and the area of the saturated object, and in the final mosaic they can
be found either in the X or in the Y direction.  On the other hand, by
exploiting  the \sext\ classification  capabilities, it was found that
only the  brightest ones are  misclassified as  galaxies, and  most of
them  are easily    deleted  by the   interactive  cleaning  procedure
described below.   Since the  automated  pipeline produces  a separate
catalog of  the  saturated stars, during the   final visual check,  we
highlight them, allowing  easy  identification and deletion  from  the
final catalogs of  their  associated cross-talk  false  detections.  A
negligible fraction of   these  objects, depending  on the   number of
saturated stars in  the mosaic, may remain after  this last check, but
are mainly classified as ``unknown'' sources.

\begin{figure}
\centering
\includegraphics[scale=0.45]{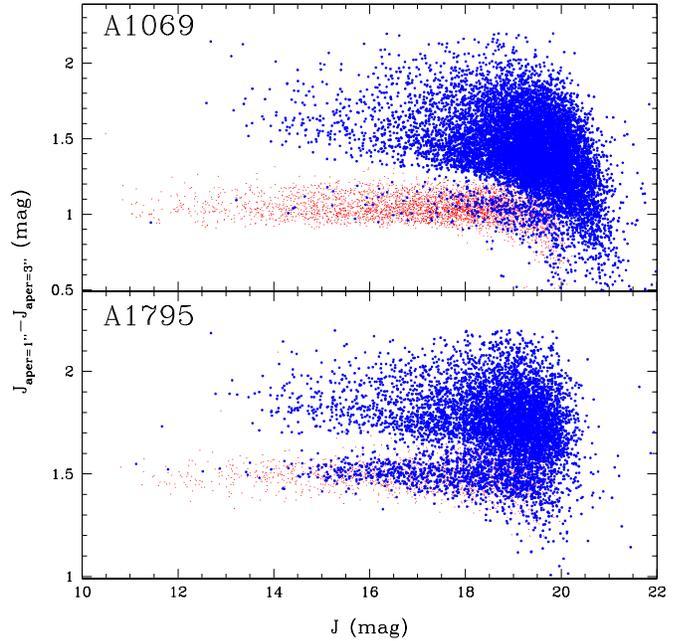}
\caption
{Plot of the difference between two  aperture magnitudes (1 arcsec - 3
arcsec diameter) vs. total magnitude for galaxies (blue bold dots) and
stars (red light dots) as  classified by \sext\ before any interactive
cleaning.  In  this diagram  the star and  galaxy loci  turn out to be
well apart down to J$\approx$19.0. In the good cases (top panel:
A1069, with  quite stable   seeing during observations,  $\tau=0.21$),
only a few galaxies are  misclassified, and even at bright magnitudes.
Alternatively,  in cases  of  strong  seeing variations (bottom  panel:
A1795, $\tau=0.32$), many more misclassifications  of galaxies may be
found (bold dots in the locus  of stars).
\label{fig:cleaning}}
\end{figure}

Since the number of objects is  too large for any individual analysis,
we purged spurious detections  and star/galaxy misclassifications with
our interactive    tool, which generates   several  plots of different
combinations of parameters, like:
\begin{itemize}
\item \mumax vs. J,K ,   where \mumax\  is  the  surface brightness   of the
  brightest pixel of an object and J,K is the total apparent magnitude
  \texttt{MAG\_AUTO}.
\item  J,K vs. $\log_{10}(Isophotal\  Area)$, where $Isophotal\ Area$ is the
area in pixels of each object at the threshold level.
\item $\log(\fwhm)$ vs. $\log_{10}(Isophotal\ Area)$.
\item J,K(~$\!\rm{d}\!\leq\!1''$)-J,K(~$\!\rm{d}\!\leq\!3''$) vs. J,K,  where
$\rm{d}$ is the aperture diameter.
\end{itemize}

In all these  diagrams, the  stars populate  a rather  narrow and well
defined  region, while galaxies are  more spread  throughout the plane
(see also, Paper-II). As  an example, in Fig.\ref{fig:cleaning} a plot
of the difference between two aperture magnitudes (1 arcsec - 3 arcsec
diameter) vs.  total \jf\ magnitude  for galaxies (blue bold dots) and
stars  (red  light   dots)   before  any  interactive   cleaning,   is
presented. It is  apparent that for the  cluster A1795, due to  seeing
variations and background conditions, a  considerable amount of objects
with
\texttt{CLASS\_STAR} typical of galaxies actually populate the region
of  stars. In the case  of A1069, the more  stable PSF causes a better
classification. Still, at faint magnitudes, a non-negligible amount of
misclassifications is left.    Thus, a further  step  turns out to  be
necessary in any case to clean the catalogs. To this aim, the outliers
and/or   misclassifications   highlighted   by   these   diagrams  are
interactively selected, and, if wanted, the  cleaning pipeline shows a
tile-mosaic of  these objects for visual   inspection to easily select
spurious  detections. Then, simple commands  allow us to look at their
location in the  original  mosaics,  list their  parameters  from  the
catalog, delete them or change their classification.

At  the   end  of  this process the   degree   of misclassification of
relatively bright sources ($J\!<\!18.5$ and $K\!<\!18.0$) is less than
$1\%$,        that       is       practically     negligible      (see
Sec.~\ref{sec:completeness}) . Of course, going to fainter and smaller
objects the regions  occupied by stars and  galaxies start to mix  up,
making the classification more and more unreliable.

The  published catalogs will  be regularly updated  to correct for any
newly   found  spurious objects  and/or  misclassifications, therefore
users are encouraged to rely upon the latest available version of the
catalogs.

%%%%%%%%%%%%%%%%%%%%%%%%%%%%%%%%%%%%%%%%%%%%%%%%%%%%%%%%%%%%%%%%%%%%%%
%%%%%%%%%%%%%%%%%%%%%% CATALOGS DESCRIPTION %%%%%%%%%%%%%%%%%%%%%%%%%%
%%%%%%%%%%%%%%%%%%%%%%%%%%%%%%%%%%%%%%%%%%%%%%%%%%%%%%%%%%%%%%%%%%%%%%

\subsection{Catalogs Description}

In Tab.\ref{tab:Catalogs} we present an example of the entries in
the near-infrared photometric   catalogs. The parameters   stored for
each object are the following (in parenthesis we give the name of the
\sext's output parameter):

\begin{itemize}
\item  ID: objects internal identification, it is unique for all
catalogs of the WINGS survey.
\item  ($\alpha_{Bary}$,$\delta_{Bary}$): equatorial coordinates (J2000.0) of the barycenter.
\item  ($\alpha_{Peak}$,$\delta_{Peak}$): equatorial coordinates (J2000.0) of the brightest pixel.
\item  Area: area at the detection threshold.
\item  $r_{Kron}$: Kron radius used to compute the \texttt{MAG\_AUTO} magnitude.
\item  \fwhm: full width at half maximum assuming a gaussian core as
calculated by \sext.
\item  b/a : axis ratio of the source.
\item  PA : position angle of the major axis (North=0$^o$, measured counter-clockwise).
\item  $\mu_{max}$ : Surface brightness of the brightest pixel.
\item  \texttt{MAG\_BEST} : \sext's best total magnitude estimate.
\item  \texttt{MAG\_AUTO} : \sext's Kron (total) aperture magnitude.
\item  \texttt{MAG(4.3kpc),   MAG(10.8kpc),  MAG(21.5kpc)}:  magnitudes  within
  apertures of   diameter 4.3kpc, 10.8kpc  and 21.5kpc,  respectively,
  measured at the   clusters' redshift.  Since we  will   adopt in 
  future  papers  of the survey  $H_0=70$  (at variance with Paper-II,
  where  $H_0=75$  has been  used), the  apertures in  kpc are slightly
  different  from those given in  Paper-II  (4/10/20 kpc): however, the
  value in arcseconds remains the same.
\item   \texttt{MAG(0\farcs7),  MAG(1\farcs0),  MAG(1\farcs6)...}: magnitudes
  measured within  8 different fixed apertures\footnote{Three of these     fixed
  apertures have been chosen to match  the multi-fiber spectroscopy of
  the WINGS fields.  1\farcs6 is the projected  diameter of the fibers
  using  the  Autofib2@WHT,  while   using the 2dF@AAT   results  in a
  diameter that varies  radially in  the field  from 2\farcs16 in  the
  center to 2\farcs0 at the  edges.} in arcsec. 
\item  \texttt{FLAG\_SEX} : \sext's \texttt{FLAG} keyword.
\item  \texttt{CLASS\_STAR} : \sext's stellarity index.
\item \texttt{WINGS\_FLAG}: summarizing flag column reporting
pipeline classification and photometric quality of the objects,
using the following prescription: 
\newline
\begin{math}
\rm{WINGS\_FLAG}=a_1+2a_2+4a_3+8a_4+16a_5+32a_6\nonumber
\end{math}
\newline
$a_1=1$ if classified as star\\
$a_2=1$ if classified as galaxy\\
$a_3=1$ if classified as unknown\\
$a_4=1$ if weakly affected by neighbouring halo\\
$a_5=1$ if strongly affected by neighbouring halo\\
$a_6=1$ if in an area of confidence $<\!\!70\%$\\
where the classification is done relying upon our custom interactive
tools, the neighbouring halo can be due to saturated stars or nearby
extended objects, and the confidence level is extracted from the
final confidence maps generated by CASU pipeline (see section \ref{stacks}). 
\end{itemize}

%%%%%%%%%%%%%%%%%%%%
%%%% Cats Table %%%%
%%%%%%%%%%%%%%%%%%%%

% PRIMA PARTE

\begin{table*}[t]
  \caption{Example of  entries  in the  on-line  version  of  J and  K
  photometric catalogs. \label{tab:Catalogs} }
\tiny 
\begin{tabular}{*{13}{c}} 
\hline \hline

   & $\alpha_{Bary}$ & $\delta_{Bary}$ & $\alpha_{Peak}$ &
$\delta_{Peak}$  &   Area  &  $r_{Kron}$  &  \fwhm   &  b/a  &  PA  &
   $\mu_{max}$ & &\\    

\rb{ID}     &  (deg)  & (deg)    & (deg)   & (deg) &
  (arcsec$^{2}$) & (\arcsec) & (\arcsec) & & (deg) &(mag arcsec$^{-2}$)
& &  \\
 
\hline 

WINGSJ103934.24-082406.9  & 159.89270 & -8.40195 & 159.89267 & -8.40193 &
      8.08 & 0.70 & 1.90 & 0.73 & 99.34 & 18.70 & \ldots & \\

WINGSJ103944.19-082410.8 &  159.93416 & -8.40298 & 159.93414 & -8.40300 &
      4.68 & 0.70 & 1.65 & 0.87 & 10.52 &  19.36 & \ldots & \\

\hline

\end{tabular}

% SECONDA PARTE

\begin{tabular}{*{12}{c}} 
   \multicolumn{12}{c}{} \\
\hline \hline

 & \texttt{MAG\_BEST}  & \texttt{MAG\_AUTO} & \texttt{MAG(4.3kpc)} & 
 \texttt{MAG(10.8kpc)} & \texttt{MAG(21.5kpc)} &\texttt{MAG(0\farcs7)} &
   \texttt{MAG(1\farcs0)} & \texttt{MAG(1\farcs6)} &
    \texttt{MAG(2\farcs0)} &   \\    

     &  (mag)  & (mag)    & (mag)   & (mag) &  (mag) & (mag) & (mag) &
     (mag ) &  (mag)  &    &   \\
 
\hline 

\ldots  & 17.63 & 17.63 & 17.69 & 17.62 &
      17.84 & 19.87 & 19.17 & 18.41 & 18.13 &   \ldots   \\

\ldots & 18.34 & 18.34 & 18.36 & 18.35 &
      18.03 & 20.50 & 19.82 & 19.07 & 18.80 &    \ldots  \\ 

\hline 

\end{tabular}

% TERZA PARTE

\begin{tabular}{*{9}{c}}
    \multicolumn{9}{c}{} \\
\hline \hline

 & \texttt{MAG(2\farcs16)} & \texttt{MAG(3\farcs0)} & \texttt{MAG(4\farcs0)} &
     \texttt{MAG(5\farcs0)} & &  & &  \\    

 &  (mag)  & (mag)    & (mag)   & (mag) & \rb{\texttt{FLAG\_SEX}}  & 
     \rb{\texttt{CLASS\_STAR}}  &\rb{\texttt{WINGS\_FLAG}} &  \\
 
\hline 

\ldots & 18.05 & 17.75 & 17.64 & 17.60 & 0 &  0.02 & 2 &  \\

\ldots & 18.71 & 18.44 & 18.31 & 18.22 & 0 & 0.01 & 2 & \\ 

\hline 

\end{tabular}
\\
{\flushleft The full table, for each cluster and filter, 
is available in electronic form at the CDS.}
\end{table*}
\normalsize
%%%%%%%%%%%%%%%%%%%%%%%%
%%%% Cats Table END %%%%
%%%%%%%%%%%%%%%%%%%%%%%%

The complete/updated list of stored parameters, together with
\sext\ configuration files, are available on the WINGS website
\texttt{http://web.oapd.inaf.it/wings/}, while the exact definition of
each parameter can be found in the \sext\ manual \citep[][]{bertin96}.
The photometric  catalogs, {in the  form of Tab.\ref{tab:Catalogs} are
available in  electronic format}.  Note that  the same  caveats and/or
qualifications  given  in   Paper-II hold for   these  photometric and
geometric parameters.

The final version of all WINGS survey catalogs will be a comprehensive
cross-matched source-list, allowing multiple criteria queries with Web
based tools and  Euro-VO   facilities, and simple  identification   of
objects which have measurements and calculated quantities belonging to
different branches of the survey (U,B,V,J,K photometry, spectroscopy).

%%%%%%%%%%%%%%%%%%%%%%%%%%%%%%%%%%%%%%%%%%%%%%%%%%%%%%%%%%%%%%%%%%%%%%
%%%%%%%%%%%%%%%%%%%%%%%%%% DATA QUALITY %%%%%%%%%%%%%%%%%%%%%%%%%%%%%%
%%%%%%%%%%%%%%%%%%%%%%%%%%%%%%%%%%%%%%%%%%%%%%%%%%%%%%%%%%%%%%%%%%%%%%

\section{Data Quality\label{sec:DataQuality}}

As already mentioned  in previous sections,  the bulk of the reduction
has been  performed  at CASU with the   custom designed pipelines  for
UKIRT-WFCAM. Hereafter we present   the results of the  quality  tests
performed by our pipeline   on source extraction and   classification,
both in real and simulated mosaics.

%%%%%%%%%%%%%%%%%%%%%%%%%%%%%%%%%%%%%%%%%%%%%%%%%%%%%%%%%%%%%%%%%%%%%%
%%%%%%%%%%%%%%%%%%%%%%%%%% COMPLETENESS %%%%%%%%%%%%%%%%%%%%%%%%%%%%%%
%%%%%%%%%%%%%%%%%%%%%%%%%%%%%%%%%%%%%%%%%%%%%%%%%%%%%%%%%%%%%%%%%%%%%%

\subsection{Completeness}\label{sec:completeness}

\begin{figure}
  \centering    \includegraphics[scale=0.45]{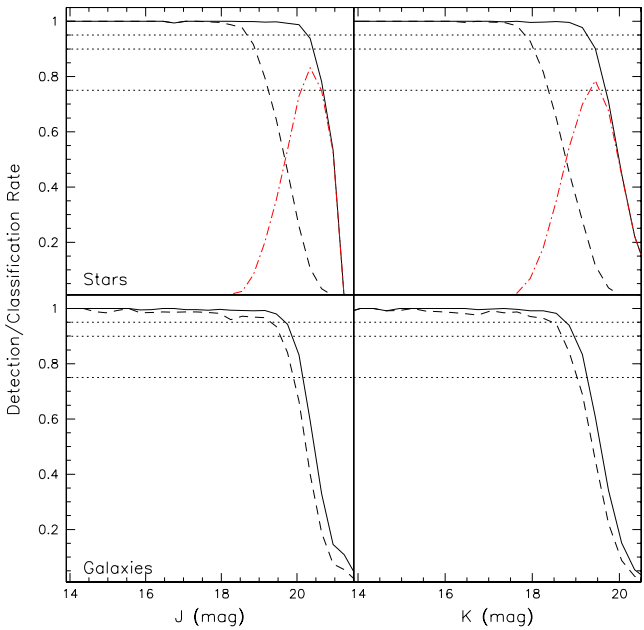}
  \caption {The global average  detection (full lines) and  successful
  classification (dashed lines) rates,  obtained from simulations  for
  the WINGS-NIR  survey in  the  J-  (left panels) and  K-band  (right
  panels).  The dotted lines  represent the 95, 90  and 75 percent  of
  completeness. The dotted-dashed  line in the upper  panels represent
  the fraction of stars misclassified as galaxies by \sext. The bottom
  panels show that, even at  bright magnitudes, some galaxies (a small
  fraction, indeed) are miss-classified as stars.   This is likely due
  to  the wide  range of  effective radii adopted  in the simulations,
  which can sometime    produce  unrealistically compact    luminosity
  profiles.  In  real mosaics,   such misclassifications  are   easily
  corrected with our interactive tool.  \label{fig:completeness}}
\end{figure}

Fig.\ref{fig:completeness}   illustrates the   global  detection   and
successful  classification  rates   (for   both stars    and galaxies)
evaluated  running \sext\  on  the  mock images,  as explained  in the
previous       Sec.~\ref{sec:stargal}         (see                also
Tab.\ref{tab:clustertab}). Full and dashed lines  are the fraction  of
the    total  input   objects,  detected   and   correctly classified,
respectively. The 90\% level of  successful classification is  reached
at J=19.0 and K=18.2  for  stars, while  the limits become  J=19.5 and
K=18.5 for extended sources.  Moreover, the 90\% detection rate limits
for     galaxies  is   reached     at     J=20.5  and    K=19.4.    In
Tab.\ref{tab:clustertab} we  report  the detection  and classification
90\% completeness limits for both  stars and galaxies for each cluster
and band, as extracted from our simulations.

It is clear that \sext\  is able to classify  correctly almost all the
galaxies  detected in simulations  (apart from  rare  and very compact
objects), while  this is not  true  for  point-like sources.  This  is
likely due to the seeing variations across the mosaics (which in a few
cases             is                      relevant,                see
Figs.\ref{fig:seeingvariation},\ref{fig:seeingrap}).                In
Fig.\ref{fig:completeness}, the three dotted horizontal lines mark the
95\%, 90\% and  75\% completeness limits of  our survey, while the red
dashed-dotted line in the upper panels represent the fraction of stars
misclassified as galaxies when relying only upon \sext's
\texttt{CLASS\_STAR}. It can be seen that for $\jf,\kf>\!\!18.0$ the
frequency of   such misclassification   becomes  relevant (see    also
Fig.\ref{fig:cleaning}) and  rapidly increases.   However, we  have to
consider  that, for faint   magnitudes, the galaxy  population becomes
gradually dominant (field galaxies) and the contribution to the galaxy
counts coming from the (usually small) fraction of misclassified stars
should be in any case negligible \citep[see, for \eg,][]{berta06}.

\begin{figure}
  \centering    \includegraphics[scale=0.45]{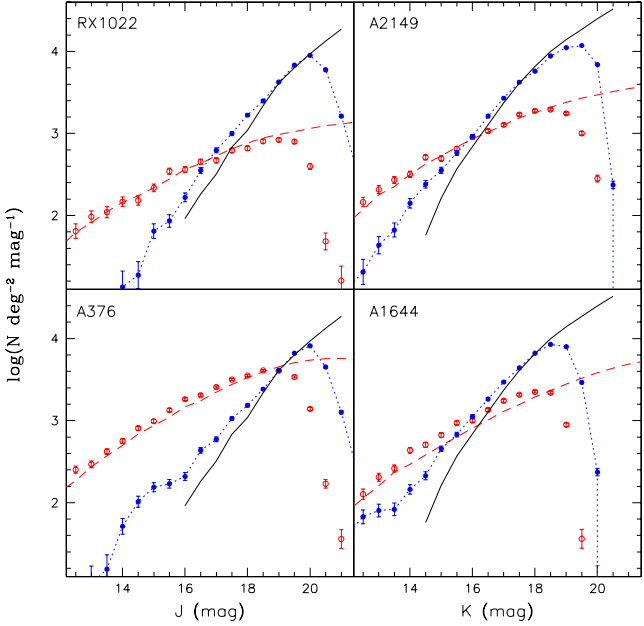}
  \caption {Number  counts in J-   (left  panels)  and K-band   (right
  panels)  compared with TRILEGAL     models of the  Galactic  stellar
  distribution  from \citet{girardi05} (red  dashed  lines)  and field
  galaxy counts from UKIDSS-UDS \citet{hartley08} (solid black lines).
  The blue full and the red open dots  are the galaxy and stars number
  counts of the  corresponding WINGS  catalogs. Poissonian error  bars
  are     mostly     of      the    dimension   of       the  symbols.
  \label{fig:numbercounts}}
\end{figure}

\begin{figure}
  \centering        \includegraphics[scale=0.45]{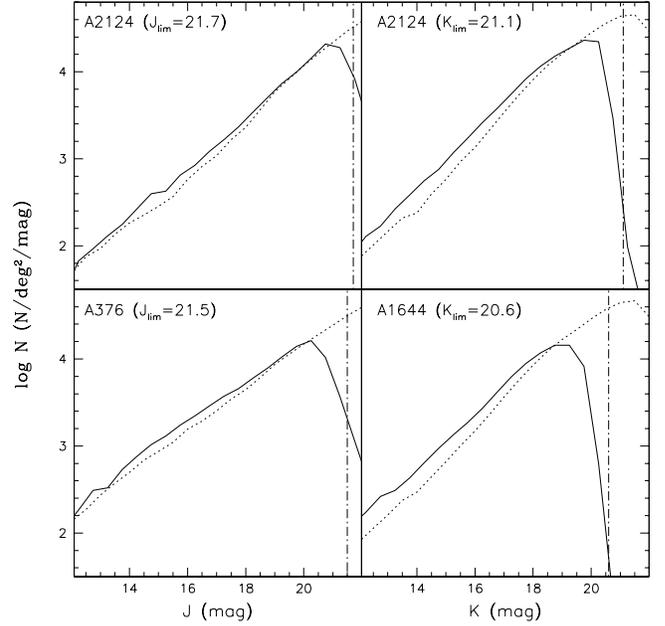}
  \caption {Magnitude counts of all detections in J- (left panels) and
  K-band   (right  panels), for  the deepest   (upper  panels) and the
  shallowest  (bottom  panels)   cluster  mosaics.   The   dash-dotted
  vertical  lines correspond  to  the estimated  theoretical magnitude
  limits from Eq.(\ref{eq:mlim}); the turnovers in the counts given in
  this  figure indicate the  detection  completeness for these cluster
  images.  Both values  are  reported in Tab.\ref{tab:catalogtab}  for
  all clusters. The dashed line is the cumulative stars+field galaxies
  distribution expected  to  be  found   in the area   of the  mosaic.
  \label{fig:limiting}}
\end{figure}

An  empirical (a posteriori)  check of completeness  can be performed,
for   each cluster in   each band,   comparing the  number  counts  of
classified objects in our catalogs, with the expected number counts of
field   galaxies  and  stars.   In Fig.\ref{fig:numbercounts}  we show
representative   cases for  J-    (left  panels)  and   K-band  (right
panels). The red dashed lines are the  expected number counts of stars
in the area of the cluster, calculated with the TRILEGAL code
\citep[][]{girardi05} in  the  photometric system of  WFCAM. The black
solid  lines are  the number counts  of  field galaxies taken from the
UKIDSS Ultra Deep    Survey \citep[see,][]{hartley08,lawrence07}.  The
open red and the full blue dots are the WINGS  number counts for stars
and galaxies, respectively, as  calculated from our source lists, with
their  Poissonian error  bars. In  general,  the agreement between the
expected  star counts and  our catalogs is outstanding, confirming the
excellent performance of the  TRILEGAL code as  a model of the Galaxy.
However, for 6  clusters out of 28,  we noticed an  excess in the real
star counts at magnitudes brighter than 18  (see for instance A1644 in
Fig.\ref{fig:numbercounts}).  Since all  these  clusters are  found to
lay approximately in the direction of the center of the Galaxy (albeit
with $\left|b\right|\geq20^o$), where a higher  fraction of halo stars
is encountered, we assume that the model  might slightly break down in
this direction.  Our galaxy counts  agree with the field number counts
only at faint magnitudes, as expected, showing the presence of cluster
galaxies at brighter magnitudes.

The excess  of   number counts  shown  in  Fig.\ref{fig:numbercounts},
revealing    the    cluster members,      can  also   be     seen   in
Fig.\ref{fig:limiting}, where the  number counts  of all the  detected
sources (full  line), is compared   with the  cumulative  (stars+field
galaxies) counts expected for the area of  the clusters (dashed line).
The approximate magnitude where the turnover in the  counts occur is a
good estimate  of the observed completeness  limit, and is reported in
Tab.\ref{tab:clustertab},   together  with  the  theoretical magnitude
limit (dotted-dashed  vertical lines)   calculated with the  following
formula:
\begin{equation}
\rm{m}_{lim} =  \rm{ZPT}-2.5\log_{10}[\nu \; (\sigma_{BG}+1) \; \rm{A_{min}} ] \label{eq:mlim}
\end{equation}
where $\nu$ is the relative  threshold cut in  units of background rms
($\sigma_{BG}$), $\rm{A}_{min}$  is the  minimum number of  contiguous
pixels required for detection  and ZPT is  the mosaic zero  point
magnitude,   normalized  to one second      exposure time and  airmass
corrected:
\begin{equation}
\rm{ZPT}=\rm{MAGZPT}+2.5\log_{10}(5)-(\chi-1)\cdot \emph{k}
\end{equation}
where $\rm{MAGZPT}$ is the  zero-point magnitude keyword found in  the
FITS header of the mosaic,  resulting from the calibration with  2MASS
performed  at CASU.  The second refers  to the WFCAM 5s exposure time,
the  term $\chi$   and  $k$ variables  are   the average  airmass  and
extinction (keywords AIRMASS and EXTINCT in FITS file), respectively.

This    theoretical limit corresponds  to  the  magnitude of an object
consisting  of $\rm{A}_{min}$ contiguous  pixels   with ADU counts  of
$\nu(\sigma_{BG}+1)$, and gives an idea  of  the overall depth of  the
mosaic imaging, since   it links photometric and  detection properties
together.  The  dash-dotted  vertical lines  in Fig.\ref{fig:limiting}
show that these  limiting magnitude values  are quite  consistent with
the faintest detected objects.

At  the final step, the  consistency among the previous three diagrams
(Figs.~\ref{fig:limiting} \ref{fig:completeness} and
\ref{fig:numbercounts}) is used as    an \textit{a posteriori}   quality
check, as well as as a good empirical  tool for possible refinement of
\sext's input parameters.  Usually, when this ideal combination
is   achieved, objects are detected  down   to the completeness  limit
without populating the source lists with unwanted spurious detections.

%%%%%%%%%%%%%%%%%%%%%%%%%%%%%%%%%%%%%%%%%%%%%%%%%%%%%%%%%%%%%%%%%%%%%%
%%%%%%%%%%%%%%%%%%%%%%%%% ASTROMETRY %%%%%%%%%%%%%%%%%%%%%%%%%%%%%%%%%
%%%%%%%%%%%%%%%%%%%%%%%%%%%%%%%%%%%%%%%%%%%%%%%%%%%%%%%%%%%%%%%%%%%%%%

\subsection{Astrometry}\label{sec:astrometry}

\begin{figure}
  \centering
  \includegraphics[scale=0.45]{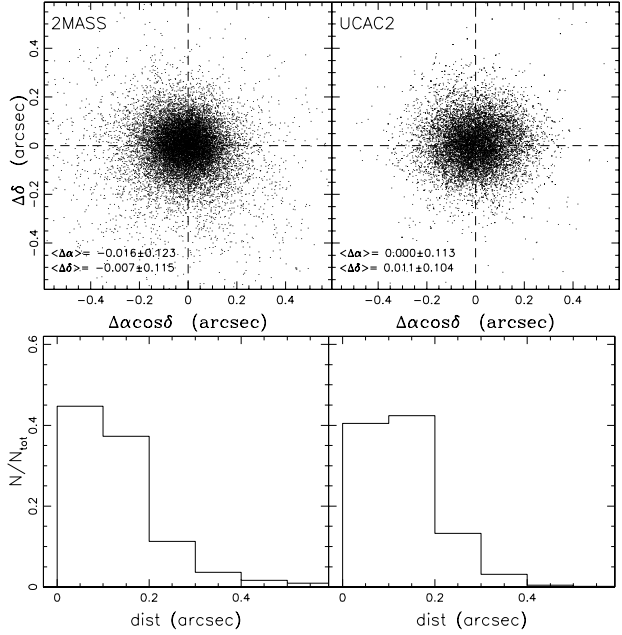}
  \caption
    {Astrometry of the WINGS-NIR survey, compared with 2MASS
    point-source and UCAC2 catalogs (upper panels) and histograms of
    the distances (bottom panels). The overall precision is of the
    order of $100\rm{mas}$ and nearly 70\% of the offsets calculated
    are below one pixel distance. The high precision astrometry of
    our survey allows safe cross-matching of catalogues extracted from
    other surveys.
  \label{fig:astrometry}}
\end{figure}

Most  of the non-linear distortion of  WFCAM over  the entire field is
accounted for, by  a cubic radial term   in the astrometric  solution.
The    CASU pipeline processes     the   raw images    considering the
differential field  distortion too, giving at  the end  an astrometric
error   usually  below  50mas\footnote{milliarcseconds.}    \citep[for
further details refer to CASU website, and][]{dye06}.   We are able to
reach that  precision only in  relative astrometry, \ie\ difference of
coordinates in two bands of the same field.  After the coadding of all
the   MEFs   into a   single  mosaic,   an additional astrometric  and
photometric re-calibration check  with   point-like sources  from  the
2MASS catalogue   is    performed.    Fig.\ref{fig:astrometry}  is   a
visualization of the astrometric precision and accuracy for our entire
cluster collection: upper panels show  the overall spread of the right
ascension and declination differences between  the positions of common
point-like sources in our catalogs and in  2MASS (left panel) or UCAC2
(right panel).  Only  sources  with  a  photometric error lower   than
0.1~mag  were considered. The  overall  zero point shift is negligible
for  all our applications, and the  rms is consistent with UKIRT-WFCAM
standard requirements, being of the order of 100mas (RMS=$112$mas).

While  the original stacks  of  each  tile  use a Zenithal  Polynomial
projection (ZPN) astrometric solution,  the final coadded  mosaics are
expressed in the standard {\em  gnomonic} tangential projection (TAN).
Thanks to  the accurate and  precise  astrometric solutions applied to
our mosaics, cross-matched source   lists in different bands  (optical
and U bands  included) are very easily obtained,  and the  fraction of
mismatches is negligible.

%%%%%%%%%%%%%%%%%%%%%%%%%%%%%%%%%%%%%%%%%%%%%%%%%%%%%%%%%%%%%%%%%%%%%%
%%%%%%%%%%%%%%%%%%%%%%%%%%% PHOTOMETRY %%%%%%%%%%%%%%%%%%%%%%%%%%%%%%%
%%%%%%%%%%%%%%%%%%%%%%%%%%%%%%%%%%%%%%%%%%%%%%%%%%%%%%%%%%%%%%%%%%%%%%

\subsection{Photometry}\label{sec:photometry}

In   Tab.\ref{tab:catalogtab} a summary    of the  properties  of  the
catalogs is  given, and the  surface brightness limits,  calculated in
the following way, are reported:
\begin{equation}
\rm{\mu}_{lim} = \rm{ZPT}-2.5\log_{10}[\nu \;
\sigma_{BG}]+2.5\log_{10}[\rm{A}''] \label{eq:mulim}
\end{equation}
where $\rm{A}''$ is the pixel area in arcseconds.  This relation gives
the  minimum surface brightness  a  pixel can  have due  to  the sigma
clipping   chosen for   the    specific    mosaic.  The  output     of
Eq.(\ref{eq:mulim}) is obviously changing from cluster to cluster, but
it  generally     settles    at        $\mu_{\rm{Jlim}}\approx22$  and
$\mu_{\rm{Klim}}\approx21$.

\begin{figure}
  \centering      \includegraphics[scale=0.45]{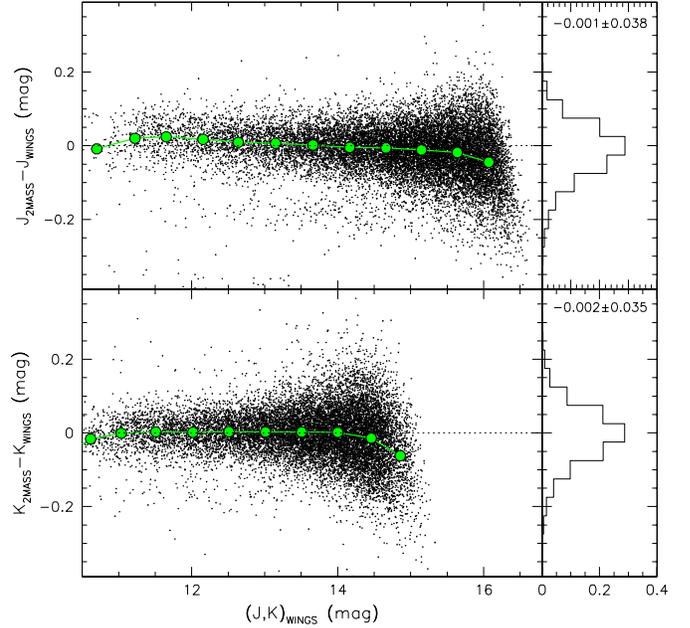}
  \caption {The  2MASS   photometry is  converted  to the  WINGS/WFCAM
  photometric  system  with  Eqs.(\ref{eq:2massj},\ref{eq:2massk}) and
  then compared with WINGS magnitudes, for the whole collection of our
  NIR  mosaics,   separately in  J- (upper   panel) and  K-band (lower
  panel).  For  a correct comparison with  2MASS we use  aperture
  ($3\farcs0$   diameter) corrected magnitudes in  the   Y axis.  The
  absence of a significant zero point shift and the relative tightness
  of the magnitude sequence demonstrates the quality of the photometry
  in the WINGS survey.  \label{fig:photometry}}
\end{figure}

Photometry  at CASU is currently based  on 2MASS, via colour equations
converting 2MASS magnitudes to the WFCAM photometric system.  The most
recently released photometric calibrations are given by
\citep[see,][]{hodgkin08}.    Neglecting  the interstellar extinction,
they are as follows:
\begin{eqnarray}
\rm{J}_{WFCAM} & = & \rm{J}_{2MASS} - 0.065 (\rm{J}_{2MASS}-\rm{H}_{2MASS}) \label{eq:2massj} \\
\rm{K}_{WFCAM} & = & \rm{K}_{2MASS} + 0.010 (\rm{J}_{2MASS}-\rm{K}_{2MASS}) \label{eq:2massk}
\end{eqnarray}
Due   to   the improved  precision   in  the  fitting algorithm, these
equations are different from those given in the early-data-release by
\citet{dye06}.  Different tests carried out at CASU suggest that the
2MASS  calibration is    delivering  photometric zero-points   at  the
$\pm2\%$  level, confirming  the   excellent results achieved by  the
2MASS survey team in ensuring a reliable all-sky accurate calibration.
While  coadding   the  single  MEFs to  obtain  the   final  mosaic, a
photometric re-calibration check based on  2MASS catalogs is performed
again,  to  assure a spatially homogeneous  zero  point throughout the
mosaic.  Fig.\ref{fig:photometry} shows  the  difference between 2MASS
and WINGS aperture  corrected magnitudes of  point-like sources (after
applying Eqs.\ref{eq:2massj} and  \ref{eq:2massk}),  vs.   WINGS total
magnitudes in the  J (upper panel) and  K (lower panel) bands, for the
complete survey.   No    significant zero point  shift  is  recovered,
confirming   the  accuracy of   the  photometric   calibration of  our
WINGS-NIR survey.
 
\begin{figure}
  \centering      \includegraphics[scale=0.45]{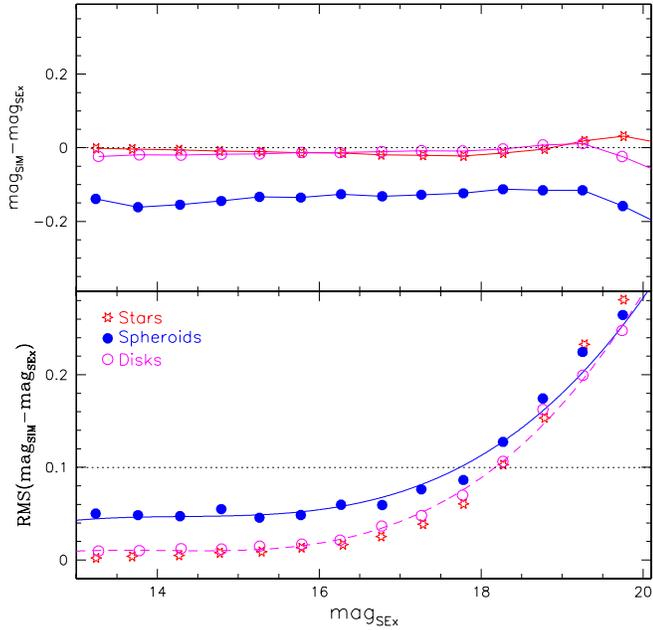}
  \caption {Global photometric precision and accuracy as obtained from
  simulations  for   stars (red  starred  dots),  disks  (open magenta
  circles)  and  spheroids (full blue   circles).  (Top panel) average
  differences between the input  and   \sext\ magnitudes for bins   of
  0.5mag of input total magnitude. (Bottom panel)  rms of the previous
  differences. While the magnitudes of  stars and disks are  perfectly
  recovered, there  is a systematic shift of  0.2mag in the spheroidal
  \dev\ magnitudes, due to the loss of light in the wings of simulated
  galaxies.  \label{fig:simulation}}
\end{figure}

Photometric errors   assigned  by \sext\   are unrealistically  small,
because they are based only on the photon-noise statistics. The use of
the so called confidence-maps (weight-maps  for \sext\ users) helps in
computing more realistic values of photometric uncertainties. However,
we   have   verified   that this latter    approach  still   leads  to
underestimates of the  photometric errors.  Therefore, we preferred to
use the    simulations  to recover  the    effective precision of  our
measurements. Nevertheless, \sext\   errors will be  available  in the
WINGS website queries forms.

In Fig.\ref{fig:simulation} we present   the global results (J  and  K
band together) of photometry checks from the  simulations.  In the top
panel the differences  between input and  output \sext\  magnitudes as
measured   from the   mock images   for    stars (red starred   dots),
exponential disks (open magenta  circles)  and spheroidal \dev\  (blue
filled circles) are plotted for 0.5~mag bins of  total magnitude.  The
agreement  is quite good, apart from  the constant shift of 0.2mag for
spheroidal galaxies  even at bright magnitudes.    This effect is well
known  from   the literature  \citep[][]{franceschini98} and   is also
recalled in Paper-II. It is caused by the loss of light from the wings
of profiles of spheroidal \dev\ galaxies in the background noise.  The
bottom panel of the same figure presents the  overall precision of our
photometry (the RMS of measurements  is plotted  for 0.5mag bins,  the
symbols are as above), with the  corresponding polynomial fit for
spheroids (blue full line) and exponential disks (magenta dashed line)
to be used as the RMS value for all magnitudes of the catalogs:
\begin{eqnarray}
\rm{RMS}_{early} & = &- 3.675908+0.778114m_{early}+\label{eq:errors1} \\
& & -0.054269m^2_{early}+0.001263m^3_{early} \nonumber\\
\rm{RMS}_{late} & = & -3.765820+0.806960m_{late}+ \label{eq:errors2} \\
& & -0.057414m^2_{late}+0.001360m^3_{late}  \nonumber
\end{eqnarray} 
where  $m_{early/late}$ can be either  the total or  the aperture
magnitude of the  corresponding object.  Due  to the way these  errors
have been  computed, the statistical  Poissonian error can be ignored.
Moreover Eq.\ref{eq:errors1} is   adequate for the errors   associated
with the  ``unknown'' objects  category, while Eq.\ref{eq:errors2} can
safely be used for point like sources.

Fig.\ref{fig:simulation} shows that the   global rms is below  0.1~mag
down  to  18.3~mag  for  stars  and   disks, and almost  18.0~mag  for
spheroids.   It  is  clear that for    J-band mosaics  this  limit can
sometimes reach  18.8~mag,   while for  the  K-band   it is  found  at
$\approx$17.5~mag, depending on  overall quality of  the image.  Given
the   overall  rates  of    successful classification    deduced  from
Fig.\ref{fig:completeness}    and Tab.\ref{tab:clustertab}, it is more
than conservative to assign  those levels of photometric errors beyond
18.0~mag and 17.5~mag for the J- and K-band, respectively.

\begin{figure}
  \centering  \includegraphics[scale=0.45]{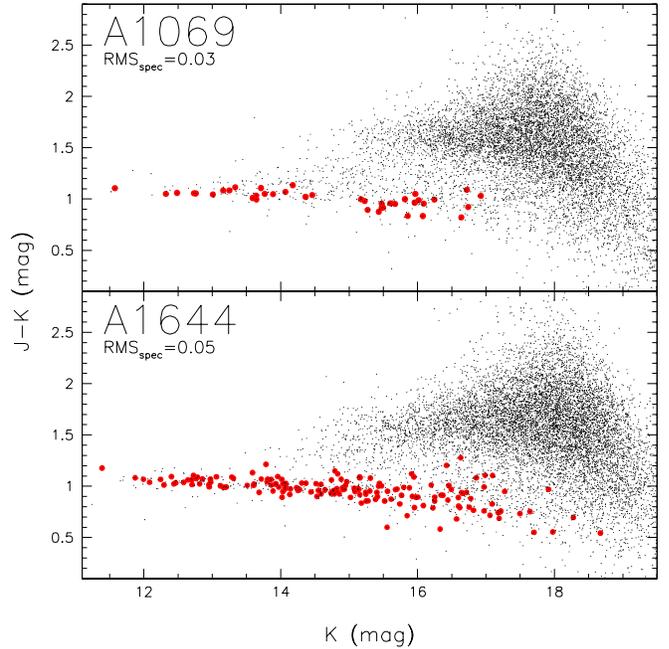}
  \caption {Color magnitude   diagrams  of classified  galaxies for  a
  selection of WINGS-NIR clusters.  The   relatively thin red  cluster
  sequences confirm the accuracy of  the photometry provided by \sext.
  The  large number of detections   obtained at fainter magnitudes are
  the  field background galaxies and   demonstrate the deepness of our
  survey.   There are also indications of  red sequences for the field
  galaxies at bright magnitudes, indicating the presence of background
  clusters  at  higher redshift.   The big  red  full  circles are the
  spectroscopically          selected  members     of   the   clusters
  \citep[][]{cava08}.  \label{fig:colormagnitude}}
\end{figure}

As an  internal   check   of   photometric   accuracy,  we  show    in
Fig.\ref{fig:colormagnitude}   two representative  examples   of total
\texttt{MAG\_AUTO}   color-magnitude diagrams  of  galaxies   from our
survey.  The  tightness  of the red sequences  (the RMS  of the
spectroscopically confirmed members  of the clusters is $0.03$mag  and
$0.05$mag) demonstrates  the internal  consistency  of the  WINGS-NIR
photometry \citep[these RMS values are found also by][]{eisenhardt07}. 

%%%%%%%%%%%%%%%%%%%%%%%%%%%%%%%%%%%%%%%%%%%%%%%%%%%%%%%%%%%%%%%%%%%%%%
%%%%%%%%%%%%%%%%%%%%%%%%%%%% SUMMARY %%%%%%%%%%%%%%%%%%%%%%%%%%%%%%%%%
%%%%%%%%%%%%%%%%%%%%%%%%%%%%%%%%%%%%%%%%%%%%%%%%%%%%%%%%%%%%%%%%%%%%%%

\section{Summary\label{sec:summary}}

In  this paper  we   have presented the    first data  release  of the
WINGS-NIR survey, comprised of J- and  K-band photometric catalogs for
a sub-sample of 28 nearby galaxy clusters  belonging to the WINGS
optical survey.  The detected  sources have been classified as stars,
galaxies and unknown objects, and for each  of them we give positions,
geometrical parameters  and  different  kinds  of total  and  aperture
magnitudes in both bands (when available).

Due to the variations of the seeing \fwhm\ over the large areas of our
mosaics and    the poor reliability   of \sext's  \texttt{CLASS\_STAR}
parameter, we used  a  specifically designed interactive  pipeline  to
improve star/galaxy   classification.   The near-infrared  WINGS data
consists   of nearly one million   detected  sources, with 150,000 and
500,000 reliably classified stars and galaxies, respectively.

This unique  collection   of data turns out  to   be 90\%  complete in
detection limits at J=20.5 and K=19.4, and in classification limits at
J=19.0 and K=18.5 (limits vary from  cluster to cluster depending
on seeing   conditions  and total integration   time),   values to be
compared with 2MASS, J=15.8 and  K=14.3 for point-like sources, UKIDSS
Large Area  Survey J=20.0  and K=18.4,  and UKIDSS Ultra   Deep Survey
J=25.0      and        K=23.0    \citep[for        further    details,
see][]{skrutskie06,dye06,lawrence07}.

Using  this new,  extensive  and comprehensive catalog,  we will study
several   different properties   and characteristics  of  low-redshift
cluster galaxies.  The near-infrared data presented in this paper will
be combined with  our optical photometry, morphology and  spectroscopy
catalogs   to  systematically study the   dependence  of these cluster
galaxy properties on their stellar mass (since this is traced by the J
and  K  bandpasses).   Furthermore, the  study   of the J,K luminosity
functions (known to  be a good tracers  of the stellar mass function),
will   allow us  to    estimate   the  distribution of  the    stellar
mass-to-light ratio as a function of the cluster-centric distance, for
different morphological types.   For the  galaxies  in our sample  for
which   we can derive  reliable  near-infrared  surface photometry and
structural parameters   (e.g.,  $R_{e}$,   $<\!\mu_e\!\!>$),  we  will
investigate the behaviour of our  sample with  respect to the  various
scaling relations   (i.e.,  Fundamental  Plane,   Kormendy  relation).
Broad-band  spectral energy  distributions  will be  generated for the
galaxies  in our  sample by  combining the   optical and near-infrared
photometry, giving  us  further  information  on stellar   content and
cluster membership. Also, using our cluster sample, we will be looking
at what effect the use of a near-infrared total cluster luminosity has
on the Fundamental Plane of Galaxy Clusters.

The final  version  of the complete WINGS   survey catalogs will be  a
comprehensive  cross-matched  source-list, allowing multiple  criteria
queries  with Web  based  tools and  Euro-VO  facilities. Through  the
website the identification   of  objects which  have  measurements and
calculated quantities from different branches of the survey (U,B,V,J,K
photometry, surface  photometry, morphology and spectroscopy)  will be
made easier by means of specific web applets.

\begin{acknowledgements}
  T.   Valentinuzzi   acknowledges a    post-doc fellowship  from  the
  Ministero  dell'Istruzione,   dell'Universit\`a  e   della   Ricerca
  (Italy).   He also  thanks for help   and  useful discussions:  Omar
  Almaini, Sebastien   Foucaud, Roberto  Caimmi,  Stefano Berta,  Luca
  Rizzi, Alessia Moretti,  and Stefano Rubele.  D. Woods  acknowledges
  the  OPTICON travel  funding  scheme  and  Australian Gemini  Office
  research  funds  both  which    made it possible  to  obtain   these
  observations at UKIRT.  He also thanks  his Padova collaborators for
  their  hospitality  and generosity   during his visit   for the team
  meeting.Marco  Riello and Tiziano  Valentinuzzi  would like to thank
  Mike Irwin for making the  stand-alone versions of the CASU pipeline
  software available and for   helpful discussions and suggestions  on
  the data processing issues related to this  work.  We also thank the
  unknown referee for useful suggestions and comments which stimulated
  discussion and resulted in an improved paper.
 	
  These  observations   have  been funded   by   the Optical  Infrared
  Coordination  network (OPTICON), a major international collaboration
  supported by the Research  Infrastructures Programme of the European
  Commission?s Sixth Framework  Programme. 

  We want to  acknowledge the Terapix  software group  for the immense
  help bestowed  by their software utilities, and  CASU  for the great
  job done with pre-reduction of images and in keeping databases.

  This  research has made use of  the NASA/IPAC Extragalactic Database
  (NED) which is operated by the Jet Propulsion Laboratory, California
  Institute of    Technology,   under  contract   with   the  National
  Aeronautics and Space Administration.

  IRAF (Image   Reduction and  Analysis   Facility)  is   written  and
  supported   by the IRAF programming  group   at the National Optical
  Astronomy Observatories (NOAO) in Tucson, Arizona.  NOAO is operated
  by the Association of Universities for Research in Astronomy (AURA),
  Inc.  under   cooperative   agreement with   the   National  Science
  Foundation.

\end{acknowledgements}

\clearpage

\renewcommand{\arraystretch}{1.2}
{\small
\clearpage
\begin{longtable}{lccccclccccc}
\caption{WINGS-NIR clusters sample with relevant useful quantities.\label{tab:clustertab}}\\
\hline
Cluster & RA & DEC & Redshift & Lx/$10^{44}$ & Pixel & RUN &
\multicolumn{2}{c}{\fwhm''}& Stacks &
\multicolumn{2}{c}{Total Area} \\
\cline{8-9}\cline{11-12}
  & hh:mm:ss & dd:mm:ss & & $\rm{erg}\,\rm{s}^{-1}$  & kpc & & min & max &   &  deg$^{2}$ & Mpc$^{2}$ \\

\hline
A119      & 10:39:43.4 &  -08:41:12.4  &0.0442     &1.65     &0.174  & 05B-K   &0.79&0.88  &16&0.781&7.66 	\\ \hline

A376    & 02:46:03.9 & +36:54:19.2   &   0.0484 &    0.71  &  0.190  & 05B-J & 1.05&1.17 & 8  &  0.777   & 9.09 \\\hline

         &           & &    &         & &05B-J&1.28&1.44          & 8 & & 		\\
\rb{A500}& \rb{04:38:52.5} & \rb{-22:06:39.0}  &\rb{0.0670}&\rb{0.72}&\rb{0.256}&05B-K&1.21&1.42& 16&\rb{0.776}&\rb{16.48} 	\\\hline

A602    & 07 53 26.6 & +29 21 34.5 &   0.0619 &    0.57  &  0.238  & 06B-K & 0.88&0.94 & 12  &  0.778   & 14.29 \\\hline

         &          & &     &        &    &05A-J&1.02&1.47         &8 &0.779& 7.47	\\
\rb{A957x}& \rb{10:13:38.3} & \rb{-00:55:31.3}   &\rb{0.0436}&\rb{0.40}&\rb{0.172}&06B-K&0.75&0.92&8 &0.782& 7.50	\\\hline

         &          & &     &          &  &06B-J     & 0.91   & 1.11       &8 &0.777  &   13.09\\
\rb{A970}& \rb{10:17:25.7} & \rb{-10:41:20.2}  &\rb{0.0587}&\rb{0.77}&\rb{0.228}&05A-K&1.06&1.26&16&0.775& 13.05	\\\hline

           & 		     & 			 &            &       & & 05A-J&0.98&1.19          &8 &0.783& 		\\ 
\rb{A1069} & \rb{10:39:43.4} &  \rb{-08:41:12.4}  & \rb{0.0650}&\rb{0.48}&\rb{0.250} & 05A-K &0.88&1.12&16&0.782&\rb{15.86} 	\\ \hline

          &  & &       &         & &06B-J    &0.80&0.93&          8 & &					\\
\rb{A1291}& \rb{11:32:23.2} &  \rb{+55:58:03.0}  &\rb{0.0527}&\rb{0.22}&\rb{0.206}&05A-K&0.99&1.15&16&\rb{0.778}&\rb{10.70} 	\\ \hline

           &      &  &     &         & &05A-J&0.89&1.15          &8 &0.781& 				\\
\rb{A1631a}& \rb{12:52:52.6} & \rb{-15:24:47.8}  &\rb{0.0462}&\rb{0.37}&\rb{0.182}&05A-K&0.82&1.10&16&0.780&\rb{8.38}	\\\hline

          &     &  &      &         & &06B-J&0.88&0.98          &8 &0.778&8.72 \\
\rb{A1644}& \rb{12:52:52.6} & \rb{-15:24:47.8}  &\rb{0.0473}&\rb{1.80}&\rb{0.186}&05A-K&0.89&1.15&16&0.772&8.65 	\\\hline

          &      &  &     &         & &05A-J&1.42&1.65          &8 &0.780&14.56	\\
\rb{A1795}& \rb{13:48:52.5} &  \rb{+26:35:34.6}  &\rb{0.0625}&\rb{5.67}&\rb{0.240}&05A-K&0.95&1.57&8 &0.387&7.22 	\\\hline

          &       & &    &      &     &06B-J &  0.72    & 0.91     &12 &0.778&  				\\
\rb{A1831}& \rb{13:59:15.1} &  \rb{+27:58:34.5}  &\rb{0.0615}&\rb{0.97}&\rb{0.238}&05A-K&0.93&1.07&16 &0.779&\rb{14.23} \\\hline

          &      & &     &          & &06B-J&0.90&1.03          &8 &0.779&7.47 \\
\rb{A1983}& \rb{14:52:55.3} &  \rb{+16:42:10.6}  &\rb{0.0436}&\rb{0.24}&\rb{0.172}&05A-K &0.79&0.98&16&0.772& 7.40	\\\hline

A1991     & 14:52:55.3     & +16:42:10.6 & 0.0587    &  0.69  & 0.228 &05A-J&0.88&1.00          &8 &0.779& 13.12	\\\hline

          &      & &     &       &  &06A-J &0.95&1.19          &8 &0.778& 6.62	\\
\rb{A2107}& \rb{15:39:39.0} &  \rb{+21:46:58.0}  &\rb{0.0412}&\rb{0.56}&\rb{0.162}&05A-K&0.85&1.03&16&0.748& 6.36	\\\hline

          &       & &        &       &   &05A-J&1.38&1.18          &8 &0.780& 16.05	\\
\rb{A2124}& \rb{15:44:59.0} &  \rb{+36:06:33.9}  &\rb{0.0656}&\rb{0.69}&\rb{0.252}&06A-K&0.80&1.03&16&0.782& 16.09	\\\hline

A2149   &16:01.28.1 & +53:56:50.4   &   0.0679 &    0.42  &  0.260   & 06A-K& 0.79&0.98 & 20  & 0.781&17.11	\\\hline

          &          & &     &        &  &06A-J&0.89&1.03          &8 &0.780& 12.91	\\
\rb{A2169}& \rb{16:13:58.1} &  \rb{+49:11:22.4}  &\rb{0.0586}&\rb{0.23}&\rb{0.226}&05A-K&0.83&0.96&16&0.771& 12.76	\\\hline

A2382   &21:51:55.6 & -15:42:21.3  &   0.0618 &    0.46  &  0.238   & 05A-K& 0.89&1.25 & 16  & 0.770&14.13	\\\hline

A2399   & 21:57:01.7 &  -07:50:22.0  &   0.0579 &    0.51  &  0.224   & 05B-K& 0.91&1.14 & 16  & 0.781&12.70 \\\hline

A2457   & 22:35:40.8 &   +01:29:05.9  &   0.0594 &    0.73  &  0.230  & 05B-K & 0.84&1.00 & 16 &  0.780   & 13.37 	\\\hline

A2572a  & 23:17:12.0 &  +18:42:04.7  &   0.0403 &    0.52  &  0.160  & 06A-K & 0.79&0.94 & 16 &  0.781   & 6.48 \\\hline

A2589   & 23:23:57.5 &  +16:46:38.3  &   0.0414 &    0.95  &  0.164  & 06A-K & 0.87&1.07 & 16 &  0.781   & 6.81 \\\hline

            &        & &       &        &   &05B-J&0.98&1.66         &8 &0.777& 9.28	\\
\rb{IIZW108}& \rb{21:13:55.9} & \rb{+02:33:55.4}  &\rb{0.0493}&\rb{1.12}&\rb{0.192}&05A-K&0.87&1.03&16&0.780& 9.32	\\\hline

          &          & &     &      &    &06B-J&0.92&1.03          &8 &0.780& 				\\
\rb{MKW3s}& \rb{15:21:51.9} & \rb{+07:42:32.1}&\rb{0.0450}&\rb{1.37}&\rb{0.176}&06A-K&0.88 &1.06       &16&0.781 & \rb{7.83}   \\\hline

           &         & &      &        &   &05B-J&0.91&1.22         &8 &0.779& 9.30	\\
\rb{RX1022}& \rb{10:22:10.0} & \rb{+38:31:23.9}  &\rb{0.0491}&\rb{0.18}&\rb{0.192}&06B-K&0.80&0.95&8 &0.782& 9.34	\\\hline

RX1740    & 17:40:32.1 & +35:38:46.1 &   0.0430 &    0.26  &  0.170  & 05A-K & 0.91&1.23 & 12  &  0.778   & 7.28 \\\hline

          &         & &      &      &     &06A-J&0.80&0.95         &8 && 					\\
\rb{Z8338}& \rb{18:11:05.2} & \rb{+49:54:33.7}  &\rb{0.0473}&\rb{0.40}&\rb{0.186}&05A-K&0.79&0.92&16&\rb{0.778}&\rb{8.72} \\\hline

\end{longtable}}
{\small
\begin{flushleft}
  RA, DEC: coordinates of the Brightest Cluster Galaxy\\
  Redshift: from NED.\\
  $Lx$: X-ray luminosity\\
  Pixel: pixel size in kpc at the given redshift.\\
  RUN: semester and passband of observation.\\
  \fwhm: minimum and maximum estimation as shown in Fig.\ref{fig:seeingvariation}.\\
  Stacks: Interleave Stacks per mosaic used.\\
  Total Area: the effective area of the mosaic.\\
\end{flushleft}
}

{\small
\clearpage
\begin{longtable}{lcccccccccc}
\caption{WINGS-NIR catalogs useful parameters. 
\label{tab:catalogtab}}\\
\hline
Cluster & DM & Band & $\rm{m}_{lim}$ & $\rm{\mu}_{thresh}$ & Cum &
\multicolumn{2}{c}{Stars} &
\multicolumn{2}{c}{Gals} & Notes\\
\cline{7-8}\cline{9-10}
  & & $mag$ & $mag$ & $mag/''^2$ & Det&Det& Clas & Det & Clas &\\

\hline

A119         &36.46 & K & 21.2 &  21.02  &  20.0  & 19.8 & 18.2 & 19.3 & 19.0 & \\\hline

A376          & 36.66 & J & 21.5 &  22.11  &   20.2 & 20.4 & 19.0 & 20.0 & 19.6 & 1\\\hline

              && J & 21.6 &  21.93  & 21.0   & 20.0 & 18.0 & 19.6 & 19.2 & 1,2	\\
\rb{A500}     &\rb{37.40} & K & 20.8 &  21.02  &  19.7  & 19.2 & 17.7 & 18.7 & 18.3 & 2,4\\\hline

A602         & 37.22 & K & 21.2 &  21.68  & 20.0   & 20.3 & 18.7 & 19.3 & 19.2 & 1,3\\\hline

             && J & 21.5 &  22.11  &  20.3  & 20.3 & 18.5 & 19.8 & 19.4 & 2,4	\\
\rb{A957x}    &\rb{36.43} & K & 21.1 &  21.49  &  20.0  & 20.0 & 18.7 & 19.7 & 19.4 & 	\\\hline

             && J & 21.8   & 22.31    & 21.0    & 20.6   & 18.6    & 20.0  & 19.5   & 2		\\
\rb{A970}    &\rb{37.10} & K & 20.9 &  21.13  & 19.7   & 19.5 & 17.7 &18.9 & 18.5 & 1,4  \\\hline

             & & J & 21.9 &  21.95 &  21.2  & 20.2 & 18.6 & 19.8 & 19.5 & 5,6	\\ 
\rb{A1069}   &\rb{37.33} & K & 20.6 &  21.32  & 19.2   & 20.3 & 18.7 & 19.7 & 19.4 & \\\hline

             & & J & 21.8 &  22.37  & 21.0   & 21.1 & 19.2 & 20.3 & 20.2 & 	\\
\rb{A1291}   &\rb{36.85} & K & 20.6 &  21.29  & 19.5   & 19.5 & 18.2 & 19.0 & 18.8 & 1,2,5\\\hline

              && J & 21.2 &  22.09  & 20.0   & 20.1 & 18.5 & 19.5 & 19.1 & 	\\
\rb{A1631a}   &\rb{36.56} & K & 20.7 &  21.02  & 19.1   & 19.7 & 18.4 & 19.3 & 18.7 & 1,3\\\hline

              && J & 21.8 &  21.98  &  20.2  & 20.7 & 19.2 & 20.0 & 19.6 & 1,4 	\\
\rb{A1644}    &\rb{36.61} & K & 20.6 &  20.90  & 19.0  & 19.5 & 18.1 & 18.9 & 18.7 & 1,4,6\\\hline

              && J & 22.0 &  22.27  & 21.5   & 20.1 & 18.0 & 19.8 & 18.0 & 2,5 	\\
\rb{A1795}    &\rb{37.24} & K & 20.8 &  21.07  & 19.0  & 19.1 & 18.0 & 18.7 & 18.5 & 2	\\\hline

              && J & 21.7 &  22.34  & 21.1 & 21.0   & 19.5  & 20.5  & 20.1    & 6	\\
\rb{A1831}    &\rb{37.20} & K & 20.5 &  21.19  & 19.3 & 19.3 & 17.5 & 19.0 & 18.6 & 1,6	\\\hline

              && J & 21.9 &  22.63  & 20.6 & 21.0 & 19.5 & 20.3 & 20.0 & 1,4	\\
\rb{A1983}    &\rb{36.43}  & K & 20.2 &  20.74  & 18.9 & 19.0 & 18.0 & 18.5 & 18.5 & 4	\\\hline

A1991         & 37.10 & J & 21.7 &  22.10  &  21.1  & 20.8 & 19.0 &20.0 & 19.5 & 	\\ \hline

              && J & 21.7 &  22.22  &  21.2  & 20.7 & 19.1 & 20.0 & 19.6 & 2	\\
\rb{A2107}    &\rb{36.30} & K & 20.8 &  21.08  & 19.2 & 19.2 & 18.0 & 19.3 & 19.1 & 1,4	\\\hline

              && J & 21.7 &  22.29  &  21.2  & 20.2 & 18.7 & 20.0 & 19.7 & 	\\
\rb{A2124}    &\rb{37.35} & K & 21.1 &  21.36  & 20.0   & 19.8 & 18.3 & 19.2 & 19.1 & 	\\\hline

A2149         & 37.43 & K & 20.9 &  21.11  &  20.0  & 19.7 & 18.3 & 19.3 & 19.0 & 6	\\\hline

              && J & 21.7 &  22.14  & 21.0   & 20.8 & 19.1 & 20.1 & 19.8 & 	\\
\rb{A2169}    &\rb{37.09} & K & 20.6 &  20.99  & 19.7   & 19.5 & 18.2 & 19.3 & 18.9 & 	\\\hline

A2382         & 37.21 & K & 20.3 &  20.57  &  19.5  & 19.2 & 17.7 & 18.8 & 18.3 & 2	\\\hline

A2399         & 37.07 & K & 20.8 &  21.04  &  19.6  & 19.5 & 18.2 & 19.2 & 18.6 & 1,4	\\\hline

A2457         & 37.12 & K & 20.9 &  21.14  &  19.8  & 19.7 & 18.0 & 19.3 & 18.7 & 4	\\\hline

A2572a        & 36.25 & K & 21.0 &  21.25  &  19.9  & 19.7 & 18.4 & 19.3 & 19.2 & 2,5	\\\hline

A2589         & 36.31 & K & 21.1 &  21.36  &  20.0  & 19.8 & 18.0 & 19.5 & 18.8 & 2,5	\\\hline

             && J & 21.5 &  21.83  & 20.4   & 20.5 & 19.0 & 19.8 & 19.0 & 1,2,4	\\
\rb{IIZW108} &\rb{36.70} & K & 21.03 &  21.32  & 19.8   & 19.8 & 18.3 & 19.5 & 19.0 & 1,4,6\\\hline

             && J & 21.9 &  22.56  & 20.5   & 21.0 & 19.6 & 20.3 & 20.0 & 1	\\
\rb{MKW3s}   &\rb{36.50} & K & 21.0 & 21.35 & 20.0   & 19.7  & 18.3  & 19.3  & 18.7  & 	2,4\\\hline

             && J & 21.5 &  22.10  &  20.2  & 20.5 & 18.9 & 19.9 & 19.6 & 	2	\\
\rb{RX1022}  &\rb{36.69} & K & 21.2 &  21.53  & 20.1   & 19.8 & 18.2 & 19.3 & 19.2 & 2	\\\hline

RX1740       & 36.40 & K & 20.6 &  20.86  & 19.1   & 19.1 & 17.8 & 18.7 & 18.2 & 1,2,4	\\\hline

             && J & 21.7 &  22.16  & 20.9   & 20.7 & 19.2 & 20.0 & 19.9 & 2		\\
\rb{Z8338}   &\rb{36.61} & K & 20.7 &  21.06  & 19.5   & 19.6 & 18.6 & 19.3 & 18.9 &	\\\hline

\hline

\end{longtable}}
{\small
\begin{flushleft}
The Cluster   column identifies  the   cluster, the DM  column  is the
cluster  distance modulus in magnitudes,  the $\rm{m}_{lim}$ column is
the    theoretical  magnitude   detection    limit calculated     with
Eq.(\ref{eq:mlim}),  $\rm{\mu}_{thresh}$ column is the minimum surface
brightness  at  the     detection  threshold   cut    calculate   with
Eq.(\ref{eq:mulim}),  the cumulative   detection column (``Cum  Det'')
lists the turnover magnitude of the number  counts diagram for all the
detected sources (see  Fig.\ref{fig:limiting}), ``Stars'' and ``Gals''
columns   represent the   average   90\% completeness   detection  and
classification   limits for   stars  and  galaxies   as  deduced  from
simulations.\\  

NOTES LEGEND:\\ 1:=  patchy and/or noisy background, 2:=
strongly  variable PSF, 3:=  many spurious  detections\\ 4:= excess of
classified stars, 5:= deficient in classified stars, 6:= exceptionally
bright star(s)
\end{flushleft}
}

\normalsize

\bibliography{wingsnir}

\end{document}